\documentclass[prd, twocolumn, lengthcheck,superscriptaddress,
  showpacs, letterpaper, nofootinbib]{revtex4-1}
\usepackage{ifpdf} \usepackage[bookmarks=false]{hyperref} \pdfoutput=1
\usepackage[usenames,dvipsnames]{color} \usepackage{graphicx}
\usepackage{graphics} \usepackage{float}

\usepackage{amsmath} \usepackage{amssymb} \usepackage{acronym}
\usepackage{multirow} \usepackage{xspace}
\usepackage{units} \usepackage{dcolumn} \usepackage{ulem}
\usepackage[bottom]{footmisc}

\def\apjl{Astrophys. J. Lett.}  % Astrophysical Journal, Letters 
 
\def\prd{Phys. Rev. D} % Physical Review D 
\newcommand{\ilya}[1]{{\color{black} #1}}

\newcommand{\partheta}{\boldsymbol{\theta}}
\newcommand{\parthetazero}{\boldsymbol{\theta_0}}

\newcommand{\ave}[2]{\left\langle #2 \right\rangle_{#1}}
\newcommand{\order}[1]{\mathcal{O}\left( #1 \right)}

\DeclareMathOperator{\cov}{cov}
\DeclareMathOperator{\erf}{erf}
\DeclareMathOperator{\se}{se}

\begin{document}
\title{Inadequacies of the Fisher Information Matrix in
  gravitational-wave parameter estimation} 
\author{Carl L. Rodriguez}
\email{cr@u.northwestern.edu} 
\author{Benjamin Farr}
\email{bfarr@u.northwestern.edu} 
\author{Will M. Farr}
\email{w-farr@northwestern.edu} 
\affiliation{Center for Interdisciplinary Exploration and Research in
  Astrophysics (CIERA) \& Dept.~of Physics and Astronomy, Northwestern
  University, 2145 Sheridan Rd, Evanston, IL 60208, USA} 

\author{Ilya Mandel} 
\email{ilyamandel@chgk.info} 
\affiliation{School of Physics and Astronomy, University of
  Birmingham, Edgbaston, Birmingham, B15 2TT}

\begin{abstract}
The Fisher Information Matrix (FIM) has been the standard
approximation to the accuracy of parameter estimation on
gravitational-wave signals from merging compact binaries due to its
ease-of-use and rapid computation time.Ê While the theoretical
failings of this method, such as the signal-to-noise ratio (SNR) limit
on the validity of the lowest-order expansion and the difficulty of
using non-Gaussian priors, are well understood, the practical
effectiveness compared to a real parameter estimation technique (e.g.,
Markov-chain Monte Carlo) remains an open question.Ê We present a
direct comparison between the FIM error estimates and the Bayesian
probability density functions produced by the parameter estimation
code \texttt{lalinference\_mcmc}.  In addition to the low-SNR issues
usually considered, we find that the FIM can greatly \emph{overestimate}
 the uncertainty in parameter estimation achievable by
the MCMC.  This was found to be a systematic effect for systems
composed of binary black holes, with the disagreement increasing with
total mass.  In some cases, the MCMC search returned standard
deviations on the marginalized posteriors that were smaller by several
orders of magnitude than the FIM estimates.  We conclude that the
predictions of the FIM do not represent the capabilities of
real gravitational-wave parameter estimation.
\end{abstract}

\maketitle
\section{Introduction}

By 2015, the first generation of gravitational wave detectors capable
of detecting astrophysical sources will become operational
\cite{AdvLIGO,AdvVIRGO,CutlerThorneReview}.  Beyond the first
detection, the science promise of these instruments lies in the
ability to estimate the parameters of the generating sources from
their data.  Much of the literature outlining the
scientific potential of advanced generation detectors, particularly with respect
to the coalescence of compact binaries, has focused on
the parameter estimation capabilities
\cite[e.g.][]{ArunPE,PoissonWill}.  These studies have informed both
the design and science goals of ground-based interferometer networks.

When discussing parameter estimation, a distinction must be made
between theoretical predictions for parameter-estimation accuracy and
the actual techniques used to measure said parameters.  For most
theoretical applications, the standard approach has been the Fisher
Information Matrix (FIM), a well-known tool from statistics that has
been adapted to gravitational-wave signal analysis
\cite{FinnDetection,CutlerFlanagan}.  Computing the FIM requires only the
partial derivatives of an analytic gravitational-wave signal model.
It can be shown that, in the strong-signal limit, the inverse of the
FIM is the variance-covariance matrix of the estimated signal
parameters.  That is, the inverse of the FIM gives a first-order
estimate of how well one could, in theory, measure the parameters of a
given system.  This estimate also corresponds to the Cramer-Rao bound on
the variance of an unbiased estimator for the parameters.
 
This is in contrast to the actual techniques used in parameter
estimation, in which one calculates the overlap of the detector output
with theoretical templates of gravitational-wave signals
\cite{spinspiral2010,nestedsampling2010}.  Since most realistic
waveform templates are generated from a high-dimensional parameter
space, the difficultly lies in adequately sampling this space.
Several studies have been performed using the LIGO-Virgo Collaboration
code, \texttt{lalinference\_mcmc}, to develop a uniform framework for
Bayesian inference on gravitational-wave data using multiple sampling
algorithms \cite[e.g.][]{Veitch2012}.  One common sampling technique,
Markov-Chain Monte Carlo (MCMC), has proved to be particularly
efficient \cite[e.g.][]{vanderSluys:2008a,vanderSluys:2008b}.

An MCMC study explores the full parameter space, with fixed data; the
FIM explores the space of data at fixed parameters.  This global MCMC
exploration of the full parameter space can, in some cases, yield highly
multi-modal, non-Gaussian posterior probability distributions which
vary substantially from the uncertainties predicted by the FIM.  Given the
effective difference between these two techniques, and the prospect of
real gravitational-wave detections and parameter estimation within the
next few years \cite{AdvLIGO,AdvVIRGO,RatesPaper,S6PE}, the question
naturally arises: how well do the theoretical estimates provided by
the Fisher Matrix formalism compare to those achievable in practice
with an MCMC search?  In this paper, we attempt to answer that
question by directly comparing the standard deviations from MCMC
searches to the estimates generated by the FIM.  We confirm the
standard assumption that the Fisher Matrix is invalid at low
signal-to-noise ratios; however, we also find that the FIM fails at
high signal strengths for systems with a total mass greater than
10$M_{\odot}$.  In those cases, the Markov-Chain Monte Carlo is
constraining parameters much more accurately than the FIM estimates.

These results robustly confirm the discrepancy first noted by Cokelaer
\cite{cokelaer} in the context of real parameter estimation.  In
several ways, this paper serves as an experimental confirmation of the
warnings given by Vallisneri \cite{Vallisneri}: that the Fisher Matrix
cannot and should not be used without several checks of internal
consistency.  In section \ref{peSection}, we review the basic analytic
setup for parameter estimation of gravitational-wave signals.  We
review the Fisher Matrix formalism in section \ref{fisherSection}, and
then describe specific details of the parameter estimation code in
\ref{MCMCSection}.  We present the primary results in section
\ref{resultsSection}.  Finally, in section \ref{breakdownSection}, we
explore various causes for the breakdown of the FIM, including the
unexpected effects of prior boundaries parameters
(\ref{priorsSection}), the breakdown of the linear-signal
approximation (\ref{vallisneriSection}), the necessity of averaging
over many noise realizations for a fair comparison
(\ref{systematicSection}), and the potential influence of hidden
estimator bias (\ref{biasSection}), explored though a
reduced-dimensionality test case.  Although we speculate on several
plausible causes, we do not yet have a fully convincing explanation
for this breakdown of the Cramer-Rao bound.  Throughout the paper, we
assume $G=c=1$, and employ the summation convention over repeated
indices.  In Appendices \ref{app:cr}, \ref{app:bd}, and \ref{app:bp}
we review the derivation of the Cramer-Rao bound, its application to
situations with a hard boundary in the data-generating distribution,
and its application to systems with strong prior constraints or
boundaries.

\section{Parameter Estimation}
\label{peSection}

We begin by introducing a Bayesian formalism for parameter estimation.
We assume that the time-domain output of a gravitational-wave detector
can be written as an additive combination of nature's gravitational
waveform $h_0$ and the noise of the detector $n$.  We further assume
that this noise is stationary and Gaussian with mean zero.  With these
assumptions the detector output is
\begin{equation}
s = n + h_0 .
\label{SignalAddition}
\end{equation}
We can write the probability of a specific data realization $s$
conditioned on the waveform parameters, $\partheta$, as a Gaussian
probability density of the residuals once the waveform $h$ has been
subtracted,
\begin{align}
  p(s | \partheta) &\propto \exp\left[-\frac{1}{2}\left<n|n
    \right>\right] \nonumber \\ &= \exp\left[-\frac{1}{2}\left < s -
    h(\partheta) | s-h(\partheta)\right >\right].
  \label{likelihood}
\end{align}
Here $\partheta$ is the set of parameters for the template waveform,
not the parameters of the actual signal, $h_0$, which we refer to as
$\parthetazero$.  The inner product, $\left< ~|~ \right> $, is
defined using the noise spectrum of the detectors as
\begin{equation}
  \left<a|b\right> \equiv 4 \Re \int df
  \frac{\tilde{a}(f)\tilde{b}^*(f)}{S_n(f)} ,
  \label{innerProduct}
\end{equation}
where $S_n(f)$ is the one-sided power spectral density of the noise as
a function of frequency, and $\tilde{a}(f)$ and $\tilde{b}(f)$ are the
Fourier transforms of the time-domain signals $a(t)$ and $b(t)$.
 
Once we have the likelihood of the detector output \eqref{likelihood}, we
employ Bayes Rule to obtain the posterior probability of the system
parameters $\partheta$ given the output $s$ as
\begin{align}
  p(\partheta | s) &= \frac{p(\partheta)p(s |
    \partheta)}{p(s)} \nonumber\\ & \propto
  p(\partheta) \exp\left[-\frac{1}{2}\big < s -
    h(\partheta) | s-h(\partheta) \big > \right] ,
  \label{posterior}
\end{align}
where $p(\partheta)$ are the prior probabilities of the source
parameters and $p(s)$ is a normalization constant.  The prior
information can come either from physical limits in the parameter space, or
from \emph{a priori} knowledge of astrophysical systems. If we pick
a set of parameters $\partheta \approx \parthetazero$ such
that $h(\partheta) \approx h_0$, then the posterior \eqref{posterior}
will be near a global maximum; however, the presence of noise will in
general deflect the maximum of the posterior away from
$\parthetazero$. That is, in the presence of noise, there is
no guarantee that the posterior is maximized at the true parameters of
the system.  The goal of parameter estimation is to sample the
available parameter space, considering all areas of posterior support,
to determine the posterior probability density function on the
parameters of the signal.  We will discuss one such sampling
technique, Markov-Chain Monte Carlo, in Section \ref{MCMCSection}.

\subsection{Fisher Information Matrix}
\label{fisherSection}
 
With the above machinery in place, the Fisher Matrix can be motivated
as follows.  Expand the template about the $\parthetazero$
waveform as
\begin{equation}
  h(\partheta) = h_0 + \Delta \theta^i h_{i} ~...
  \label{expansion}
\end{equation}
where the $h_{i}$ are the partial derivates with respect to the
$i^{th}$ parameter, evaluated at the true values, and $\Delta\theta^i
= \theta^i - \theta_0^i$.  Truncation at first-order in the partial
derivatives is sometimes referred to as the linearized-signal
approximation (LSA).  Inserting equations \eqref{expansion} and
\eqref{SignalAddition} into the posterior probability
\eqref{posterior} yields the LSA posterior probability distribution:
\begin{align}
 p(\partheta | s) \propto
 p(\partheta&)\exp\Bigg[-\frac{1}{2}\left< n| n\right >
   + \label{linearizedPosterior}\\ & \Delta \theta^k \left < n | h_{k}
   \right > - \frac{1}{2} \Delta \theta^i \Delta \theta^j\left <
   h_{i}| h_{j}\right>\Bigg]\nonumber
\end{align} 
Note that the posterior is a probability distribution over parameters
$\partheta$ conditioned on the detector output, $s$.
\citet[\S IIE]{Vallisneri} shows that the LSA is equivalent to the
leading term of the posterior expanded as a series in $\epsilon
\equiv 1/\mathrm{SNR}$.  It is for this reason that we treat the
LSA as applicable in the high-SNR regime.

Calculations with Eq.~\eqref{linearizedPosterior} are simplest when
using ``flat'' priors, $p(\partheta) \sim \mathrm{const}$.  Even if
the prior is not strictly independent of $\partheta$, if the scale on
which the prior changes is much larger than the scale over which the
posterior varies, the prior can be approximated as a constant.  Under
the flat-prior assumption, the mean of $\Delta \theta^i$ is 
\begin{equation}
  \label{eq:bayesian-dtheta-mean}
  \left \langle \Delta \theta^i \right \rangle \equiv \frac{\int
    d\partheta\, \Delta \theta^i p(\partheta | s)}{\int d\partheta \,
    p(\partheta | s) } = \left( \left\langle h_i | h_j
  \right\rangle \right)^{-1} \left \langle n |  h_j \right \rangle
\end{equation}
and the covariance matrix of $\Delta \theta^i$ is 
\begin{multline}
  \label{eq:bayesian-dtheta-cov}
  \cov\left( \Delta \theta \right) \equiv \left \langle \left( \Delta
  \theta^i - \left\langle \Delta \theta^i \right\rangle \right) \left(
  \Delta \theta^j - \left \langle \Delta \theta^j \right \rangle
  \right) \right \rangle \\ = \left( \left\langle h_i | h_j \right\rangle
  \right)^{-1}
\end{multline}
These quantities are both formally conditioned on the data 
realization (i.e.\ the particular noise in $s$), though the
covariance of $\Delta \theta$ is independent of $n$ in the LSA.

Instead of using the Bayesian framework in
Eq.~\eqref{linearizedPosterior}, suppose we employ a
maximum-likelihood estimator for the parameters $\Delta \theta^i$
under the LSA.  Note that the likelihood is a distribution on the
\emph{data} conditioned on the \emph{parameters}, so this is in some
sense a reversal of the viewpoint in Eq.~\eqref{linearizedPosterior}.
The parameters that maximize the likelihood for fixed data are
\begin{equation}
  \Delta \theta^i_\mathrm{ML}(n) = \left( \left\langle h_i | h_j
  \right\rangle \right)^{-1} \left \langle n |  h_j \right \rangle
\end{equation}
\ilya{Thus, under the LSA and a flat prior, the mean lies at the peak (mode) of the posterior.}
The expectation of the maximum likelihood estimator for $\Delta
\theta^i$ over many data realizations with fixed parameters is
\begin{equation}
  \label{eq:freq-dtheta-mean}
  \left \langle \Delta \theta^i_\mathrm{ML}(n) \right \rangle_n = \left\langle
  \left( \left\langle h_i | h_j \right\rangle \right)^{-1} \left
  \langle n | h_j \right \rangle \right\rangle_n = 0,
\end{equation}
where we have used the notation $\left \langle \right \rangle_n$ to
emphasize that this expectation is taken over the distribution of
signals (i.e.\ noise realizations) with fixed parameters, in contrast
to the angle brackets in Eqs.~\eqref{eq:bayesian-dtheta-mean} and
\eqref{eq:bayesian-dtheta-cov} which are taken over parameters with
fixed data.  Note that the expected value of the maximum-likelihood
estimator for $\Delta \theta^i$ implies that the expected value of the
maximum-likelihood estimator for $\partheta$ is
$\parthetazero$, \ilya{so this estimator is unbiased}.

The covariance of the maximum-likelihood estimator for $\Delta \theta$
in the LSA is
\begin{equation}
  \label{eq:freq-dtheta-cov}
  \cov_n(\Delta\theta_\mathrm{ML}) = \left( \left\langle h_i | h_j \right\rangle
  \right)^{-1},
\end{equation}
where again we have used a subscript to indicate that the average here
is over detector outputs at fixed parameters (i.e.\ noise realization).
Comparing to Eq.~\eqref{eq:bayesian-dtheta-cov}, we see that the
covariance of the maximum-likelihood estimator (under the distribution
of noise) and the covariance of the parameters (under the Bayesian
posterior with flat priors) are equal in the LSA (this equality is
discussed at length in \citet{Vallisneri}).  

The quantity $\left\langle h_i | h_j \right\rangle$ is also the Fisher
information matrix (FIM) for the likelihood in Eq.~\eqref{likelihood}:

\begin{equation}
  \label{FIM}
  F_{ij} \ilya{ \equiv 
   -\left<\partial_i \partial_j  \log p(s|\partheta)\right >_n \Large |_{\partheta=\parthetazero} } = 
   \left\langle h_i | h_j \right
  \rangle,
\end{equation}
where we used Eq.~\eqref{likelihood}, noting that the average over different
noise realizations implies $\left< \left<a| n\right>\left<n |
b\right>\right>_n = \left<a| b\right>$, and ignoring the contribution
from the prior.   The definition of the FIM does not depend on the LSA, but under the
LSA we see that the covariance\ilya{s of both} the maximum-likelihood estimator and
the parameters under the posterior are equal to the inverse of
the FIM.

When we consider the exponential form of \eqref{linearizedPosterior}
and that the $\Delta\theta^i$ are the displacements of the waveform
parameters from the best-fit values of $\parthetazero$, we can
then treat \eqref{linearizedPosterior} as a multidimensional Gaussian
with variance-covariance $\Sigma^{ij} = (F^{-1})^{ij}$.  The standard
deviations and cross-correlations of parameters are given by
\begin{align}
  \sigma_i &= \sqrt{\Sigma^{ii}} \label{standardDeviations}
  \\ \text{cov}(\theta^i, \theta^j) &=
  \frac{\Sigma^{ij}}{\sqrt{\Sigma^{ii}\Sigma^{jj}}} \label{crossCorrelations}
\end{align}

\subsubsection{Fisher Matrix as the Cramer-Rao Bound}
 
One must be careful when discussing the full interpretation of the
Fisher information matrix, as there are two separate statistical
meanings of $F_{ij}$.  The first, outlined above, is that in the
high-SNR/LSA limit the FIM represents the inverse of the
variance-covariance matrix.  Under this interpretation, one assumes
that for sufficiently loud waveforms, the posterior \eqref{posterior}
becomes a true Gaussian, and that $\Sigma^{ij}$ describes the
uncertainties associated with the posterior for any fixed data 
realization\footnote{But note that there is an additional
  ``uncertainty'' in the posterior due to the displacement of the peak
  from the true parameters, $\parthetazero$, which is given in
  Eq.~\eqref{eq:bayesian-dtheta-mean}.  Under the LSA, this
  displacement has zero expectation under repeated data realizations
  and covariance equal to $\Sigma^{ij}$.}.  In this limit, we expect
that the uncertainties returned by parameter estimation will coincide
with those predicted by the FIM.
 
However, there is a second, equally valid interpretation of $F_{ij}$
that is frequently employed: the inverse Fisher matrix gives the
Cramer-Rao bound on the expected variance of any unbiased estimator
for $\parthetazero$ over repeated measurements at fixed
$\parthetazero$ (i.e.\ averaged over noise realizations).  Recall that
the Cramer-Rao bound is given by \cite{Jaynes,Vallisneri}
\begin{equation}
  \left | \Sigma^{ij} 
   \right| \geq \left| \Sigma^{ij}_{CR} \right|
  \equiv \left| \ilya{F_{ij}}^{-1} \right|=\left<h_i | h_j\right>^{-1}.
  \label{cramerrao}
\end{equation}
We give a derivation of the one-dimensional Cramer-Rao bound in
Appendix \ref{app:cr}.  Equation \eqref{cramerrao} is the same form of
the covariance matrix that was stated in equations
\eqref{eq:bayesian-dtheta-cov} and \eqref{eq:freq-dtheta-cov}.
However, the key difference is that, through the lens of the
Cramer-Rao bound, \eqref{cramerrao} is now the lower bound on the
covariance of unbiased parameter estimates that can be measured given
the waveform $h_0$, not the estimated errors that can be obtained in
the high-SNR/LSA limit.  We remind the reader that this implies the
standard deviations obtained from the FIM are the \textit{lower limit}
on what is achievable on average from an unbiased estimator.

A subtle point is that the Cramer-Rao bound does not actually bound
the variance of the Bayesian posterior in a trivial way, as we
shall now describe.  Recall that the Cramer-Rao bound applies to an
unbiased \emph{estimator} of the signal parameters, $\parthetazero$,
under repeated data realizations.  Choose as an estimator for
$\parthetazero$ a fair draw from the posterior, $\hat{\partheta} \sim
p(\partheta | s)$.  We will assume for the moment that the posterior
is an unbiased estimator for $\parthetazero$; this is certainly true
under the LSA (see Eq.\eqref{eq:bayesian-dtheta-mean}).  The mean of
this estimator over many data realizations with fixed
$\parthetazero$ is the mean over $\partheta$ and data realizations
of the posterior:
\begin{equation}
\left \langle \hat{\partheta} \right\rangle_n = \int ds \, \int d\partheta \, \partheta p\left(\partheta | s\right) p\left( s | \parthetazero \right).
\end{equation}
The covariance of this estimator is
\begin{multline}
  \label{eq:cov-n}
  \cov_n \hat{\partheta} = \int ds \, \int d\partheta \,
  \left(\theta^i - \left\langle
  \hat{\theta}^i\right\rangle_n\right)\left(\theta^j - \left\langle
  \hat{\theta}^j\right\rangle_n\right) \\ \times p\left(\partheta|s\right)
  p\left(s | \parthetazero\right).
\end{multline}
A bit of algebra reveals that two terms contribute to the covariance
of $\hat{\partheta}$\footnote{Note the distinction between $\cov$ and
  $\cov_n$.  The $\cov_n$ operator is defined in Eq.~\eqref{eq:cov-n},
  and involves an integral over the noise (i.e.\ data) distribution.
  The $\cov$ operator involves an integral over the posterior
  distribution for the parameter $\partheta$, and is defined by
  \begin{equation}
    \cov \partheta \equiv \int d\partheta\, \left( \theta^i - \left\langle \theta^i\right\rangle \right) \left( \theta^j - \left\langle \theta^j \right\rangle \right) p(\partheta | s),
  \end{equation}
  with 
  \begin{equation}
    \left\langle \partheta \right\rangle \equiv \int d\partheta\,
    \partheta p(\partheta | s).
  \end{equation}}:
\begin{equation}
  \cov_n \hat{\partheta} = \cov_n \left( \left\langle \partheta
  \right\rangle \right) + \left\langle \cov \partheta \right\rangle_n.
  \label{eq:WillMagic}
\end{equation}
The first term reflects the covariance under repeated data 
realizations of the posterior mean, while the second is the average
under repeated data realizations of the posterior covariance.
Informally, the former accounts for the shifting of the posterior
peak, while the latter accounts for the typical posterior width.
Together, they must satisfy the Cramer-Rao bound:
\begin{equation}
  \label{eq:CRB-posterior-draw}
  \cov_n \hat{\partheta} = \cov_n \left( \left\langle \partheta
  \right\rangle \right) + \left\langle \cov \partheta \right\rangle_n
  \geq \Sigma^{ij}_{CR}.
\end{equation}

Another estimator for the true parameters, $\parthetazero$, is the
posterior mean, $\hat{\partheta} = \left\langle \partheta
\right\rangle$.  If a draw from the posterior is an unbiased estimator
for $\parthetazero$ then so is the posterior mean.  The Cramer-Rao
bound for this estimator therefore implies
\begin{equation}
  \label{eq:CRB-posterior-mean}
  \cov_n \left( \left\langle \partheta \right\rangle \right) \geq \Sigma^{ij}_{CR}.
\end{equation}
Because covariance matrices are positive definite\footnote{Recall that
  the matrix statement 
  \begin{equation}
    \mathbf{A} \geq \mathbf{B}
  \end{equation}
  is to be interpreted as 
  \begin{equation}
    \forall \mathbf{v}, \mathbf{v}^T \mathbf{A} \mathbf{v} \geq \mathbf{v}^T B \mathbf{v},
  \end{equation}
  or, in other words, that $\mathbf{A} - \mathbf{B}$ is positive
  semi-definite.}, Eq.~\eqref{eq:CRB-posterior-mean} implies
Eq.~\eqref{eq:CRB-posterior-draw} independently of the value of
$\left\langle \cov \partheta \right\rangle_n$, so
Eq.~\eqref{eq:CRB-posterior-draw} does not constrain the posterior
variance.

Though we have not proven that there is \emph{no} bound on the mean
posterior variance under the noise distribution induced by the
Cramer-Rao bound, the argument above is suggestive that the Cramer-Rao
bound does not apply to the variance of the posterior in a trivial
way.  Of course, since the posterior mean must have a variance under
repeated signal realizations that is greater than the inverse Fisher
matrix, in a consistent analysis the posterior variance should be of
the same order on average.  We will give several examples throughout
the paper where the posterior variance is smaller than the inverse
Fisher matrix.

In the above, we have been ignoring the effect of the prior,
$p(\partheta)$.  For parameters that are tightly-constrained by the
likelihood this is justified; however, for likelihoods that are wide
enough that the prior changes appreciably over the region of
significant likelihood support or that approach a hard boundary in the
prior, the approach above is invalid.  In general, an approach to a
prior boundary introduces bias in estimators of $\partheta$ derived
from the posterior, so the unbiased Cramer-Rao bound no longer need
apply.  Several of the examples we show below where the full analysis
significantly betters the Cramer-Rao bound have parameters that place
significant posterior support near a prior boundary.  In Appendix
\ref{app:bp}, we use a toy example to illustrate the effect of hard
boundaries in the prior on the Cramer-Rao bound.

When computing \eqref{FIM}, it is important to carefully account for
potential numerical errors, both from the numerical derivatives and
from the matrix inversion required to produce \eqref{FIM}.  Our
implementation and its numerical checks are described in section III C
of \cite{RodriguezIMRI}.  Our numerical derivatives in \eqref{FIM} are
computed with a 8th order finite difference scheme with an adaptive
step size designed to minimize numeric error.  Furthermore, these
results were checked (for the intrinsic parameters) against an
analytic computation of the derivatives, and successfully reproduce
many well known results in the literature
\cite{PoissonWill,ArunPE,Vallisneri}.

The inversion of \eqref{FIM} can pose a problem if the Fisher matrices
are ill-conditioned (that is, their determinant is sufficiently close
to zero that rounding errors become evident in the matrix inversion).
However, we note here that the condition numbers (the ratio of the
largest to smallest eigenvalues of $F$) of the majority (97\%) of the
results quoted here are below the inversion limit, which is
approximately $10^{15}$ for a 64-bit infrastructure.  In addition to
this, we studied the stability of the LU inversion of the FIM with the
largest condition number, ($7.7 \times 10^{15}$), by perturbing the
$\Gamma$ entries at the decimal point of the largest derivative error
and observing the effect on the inverse.  In that case, the difference
in the entries of the variance-covariance matrix were found to be
negligible.  Furthermore, the low-dimensional tests quoted in Table
\ref{2D3DRerunsTable} were inverted with a smallest condition number
of $2\times10^{7}$, several orders of magnitude below the regime of
numerical error.  To summarize, we are confident in the full FIM
results presented here.
  
\subsection{Parameter estimation via Markov-Chain Monte Carlo}
\label{MCMCSection}
  
We use a Bayesian parameter estimation code,
\texttt{lalinference\_mcmc}, which is an enhancement of the previously
described MCMC parameter estimation code \texttt{SpinSpiral}
\cite{spinspiral2009, spinspiral2010}.  It is designed to record a
chain of samples whose distribution is
$p\left(\partheta|s\right)$.  The basic description of the
Markov-Chain Monte Carlo via the Metropolis-Hastings algorithm is as
follows \cite{Gilks99}:
  
\begin{enumerate}
\item Pick an initial point in the parameter space
  ($\boldsymbol{\theta_{\text{old}}}$), and then propose a random
  ``jump'' to a new set of waveform parameters,
  $\boldsymbol{\theta_{\text{new}}}$.  The jump follows the
  (conditional) jump probability distribution $q\left(
  \boldsymbol{\theta_{\text{new}}} | \boldsymbol{\theta_{\text{old}}}
  \right)$.
\item Calculate the posterior probability,
  $p(\boldsymbol{\theta_{\text{new}}}|s)$, of the new parameters using
  \eqref{likelihood} and \eqref{posterior}.
\item Accept the new parameters with probability
  \begin{equation}
    p_\mathrm{accept} = \min \left[ 1,
      \frac{p(\boldsymbol{\theta_{\text{new}}}|s)
        q\left(\boldsymbol{\theta_{\text{old}}} |
        \boldsymbol{\theta_{\text{new}}}
        \right)}{p(\boldsymbol{\theta_{\text{old}}}|s)
        q\left(\boldsymbol{\theta_{\text{new}}} |
        \boldsymbol{\theta_{\text{old}}} \right)} \right].
  \end{equation}
  If the new parameters are accepted, record
  $\boldsymbol{\theta_\text{new}}$ and repeat with
  $\boldsymbol{\theta_\text{old}} \gets
  \boldsymbol{\theta_\text{new}}$; otherwise, record
  $\boldsymbol{\theta_\text{old}}$, and repeat.
\end{enumerate} 
  
Depending on the jump proposal distribution, $q$, the convergence
(mixing) of the Markov chain may be rapid or slow.  We employ multiple
optimization techniques, including both specially-crafted $q$ and
parallel tempering, to ensure adequate mixing of the Markov Chains
throughout the parameter space.  The details of the algorithm can be
found in \cite{Sluys08,spinspiral2009, spinspiral2010}.
  
The MCMC has been subjected to a series of tests to validate
posterior estimates returned by the code. The estimates of several
15-dimensional analytic functions, including unimodal and bimodal
correlated Gaussian distributions and Rosenbrock functions, have
been tested against analytic functions using the Kolmogorov--Smirnov
(KS) test. The same set of two hundred injections used for this study
(generated from the prior distibrution) was also used to verify that
the estimated Bayesian credible intervals correspond to the appropriate
frequentist confidence intervals. This was done by calculating the quantile
value at the true location for each parameter in each injection in the
set. The distribution of quantiles for each parameter was then tested
for uniformity, also using KS tests.

\subsection{Signal Model}
\label{waveformSection}
  
We use a frequency domain waveform accurate to $2^{nd}$
post-Newtonian (pN) order in phase.  We restrict ourselves to
quasi-circular waveforms as a simplifying assumption.  The standard
form of the waveform model, known as the TaylorF2
approximant, is calculated via the stationary phase approximation
where the amplitude terms are truncated to leading order in frequency
\cite{BuonannoWaveform}.  In this setup, the gravitational-wave
amplitude is given by
\begin{equation}
\tilde{h}(f) = A f^{-7/6}e^{i \psi(f)},
\label{amplitude}
\end{equation}
where $A \propto \mathcal{M}_c^{5/6}\Theta(\text{angle})/D$, $D$ is
the luminosity distance of the binary, and $\psi(f)$ is the pN phase.
$\Theta(\text{angle})$ is a function of the orientation of the binary with
respect to the detector network in terms of the sky position, orbital
inclination, and the wave polarization.  In addition to the binary
component masses $m_1$ and $m_2$, it is convenient to work with the
total mass, $M\equiv m_1+m_2$, the symmetric mass ratio
$\eta$, and the chirp mass $\mathcal{M}_c$, defined by
\begin{equation}
  \eta\equiv m_1m_2/M^2~~~~\text{and}~~~~\mathcal{M}_c = \eta^{3/5}M.
\end{equation}  
Then, in terms of the Newtonian orbital velocity $v=(\pi M f)^{1/3}$,
the 2pN phase is
\begin{equation}
\psi(f) = 2 \pi f t_c - \phi_0 + \frac{\pi}{4} + \frac{3}{128
  \eta}v^{-5}\sum^{4}_{k=0}\alpha_{k}v^k
\label{phase}
\end{equation}
with coefficients
\begin{align}
\alpha_0 &= 1\\ \alpha_1 &= 0 \nonumber\\ \alpha_2 &=
\frac{20}{9}\left(\frac{743}{336}+\frac{11}{4}\eta\right)\nonumber\\ \alpha_3
&= -16\pi \nonumber\\ \alpha_4 &=
10\left(\frac{3058673}{1016064}+\frac{5249}{1008}\eta+\frac{617}{144}\eta^2
\right) . \nonumber
\end{align}
The terms $t_c$ and $\phi_0$ in equation (\ref{phase}) are constants
of integration, referring to the time and phase at coalescence,
respectively.  Although generally uninteresting physically, they must
be accounted for in any parameter estimation study of the waveform
phase.

For the TaylorF2 approximant, the standard amplitude is given
in equation \eqref{amplitude}.  However, the LALInspiral package included in the \emph{LSC Algorithm
  Library} \cite{LALcite}, which generated the waveforms for our
Markov-Chain Monte Carlo analysis, uses a non-standard definition of
the amplitude.  This causes a slight difference in the amplitude at
high frequencies compared to the standard TaylorF2 waveform.
This issue was also noted, but not accounted for, in the previous
study by Coeklear \cite{cokelaer}.  We correct this discrepancy by
using the same waveform for both the MCMC and FIM analyses.  This was
done purely for consistency, as in practice the waveform amplitude has
a minimal effect on parameter estimation for non-spinning systems.

To perform the integral defined in \eqref{innerProduct}, we used an
analytic noise curve roughly representative of initial LIGO
sensitivity, provided in \cite{DarmourNoiseCurve}, which takes the
form:
\begin{align}
 S_h(f)=&S_0\Big[ \left(\frac{4.49 f}{f_0} \right)^{-56} +
   \nonumber\\ &\left(\frac{0.16f}{f_0}\right)
   ^{-4.52}+0.52+0.32\left( \frac{f}{f_0}\right) ^2\Big],
 \label{PSD}
\end{align}
where $f_0 = 150~\text{Hz}$, and $S_0 = 9\times
10^{-46}\text{Hz}^{-1}$. We consider the complete initial detector
network consisting of the two LIGO sites (in Hanford, WA and
Livingston, LA) and the Virgo site (in Pisa, Italy), although for simplicity
we use the Initial LIGO sensitivity for all three detectors.  For a
multi-detector network, the likelihood $p(s|\partheta)$ is a product
of the likelihoods $p(s_{det}|\partheta)$ over individual detectors,
allowing us to use the above formalism with minimal modification.  We
integrate the inner product from a lower frequency cutoff of
$40\text{Hz}$ to the innermost-stable-circular orbit of the system in
question, which for a non-spinning binary is given by
\begin{equation}
  \pi f_{\text{ISCO}} = \frac{1}{6^{3/2}M}.
  \label{ISCOFrequency}
\end{equation}
   
In addition to the intrinsic parameters ($\mathcal{M}_c,
\eta$) and phasing parameters ($\phi_0,t_c$) listed above, we include an additional
5 extrinsic parameters in the waveform model in order to
explore how a complete study on real data, including
parameters such as sky location, will compare to their FIM
counterparts.  This leads to a 9-dimensional parameter space for
non-spinning systems:
\begin{equation}
\partheta = (\mathcal{M}_c, \eta,
\phi_0,t_c,D,\iota,\psi,\alpha,\delta)
\label{parameterspace}
\end{equation}
where $D$ is the luminosity distance to the binary, $\iota$ is the
orbital inclination, $\psi$ is the gravitational-wave polarization, and
$\alpha$ and $\delta$ are the right ascension and declination of the
source on the sky.

The optimal matched-filter signal-to-noise ratio (SNR) of a
gravitational wave in a single detector is
\begin{equation}
  \rho \equiv \frac{4}{\sigma} \int^{\infty}_{0}df\frac{|
    \tilde{s}(f)\tilde{h}^{*}(f)|}{S_{n}(f)}
  \label{formalSNR}
\end{equation}
where $\rho$ is the SNR and $\tilde s(f)$ and $\tilde{h}(f)$ are
the frequency-domain signal and template, respectively.  The SNRs for
multiple detectors add in quadrature.  The normalization $\sigma$,
corresponding to the standard deviation of the matched filter output
when applied to noise alone, is given by
\begin{equation}
  \sigma^2 = 4\int^{\infty}_{0} df\frac{| \tilde{h}(f)|^2}{S_n(f)}.
  \label{SNRnorm}
\end{equation}
For the MCMC analysis, we compute the SNR directly using
\eqref{formalSNR}.  In the case of the FIM, we approximate the SNR
using $\sigma$ from equation (\ref{SNRnorm}) for a network of
detectors as
\begin{equation}
\rho = \sqrt{\sum_i \sigma_i^2}
\label{snr}
\end{equation}
where the index $i$ refers to the data of the
$i^{\text{th}}$ detector.  It should be noted that equations
\eqref{formalSNR}--\eqref{snr} are only valid in case of stationary
Gaussian noise.
   
\section{Results}
\label{resultsSection} 

\begin{figure*}[htp!]
  \centering
  \includegraphics[scale=0.72]{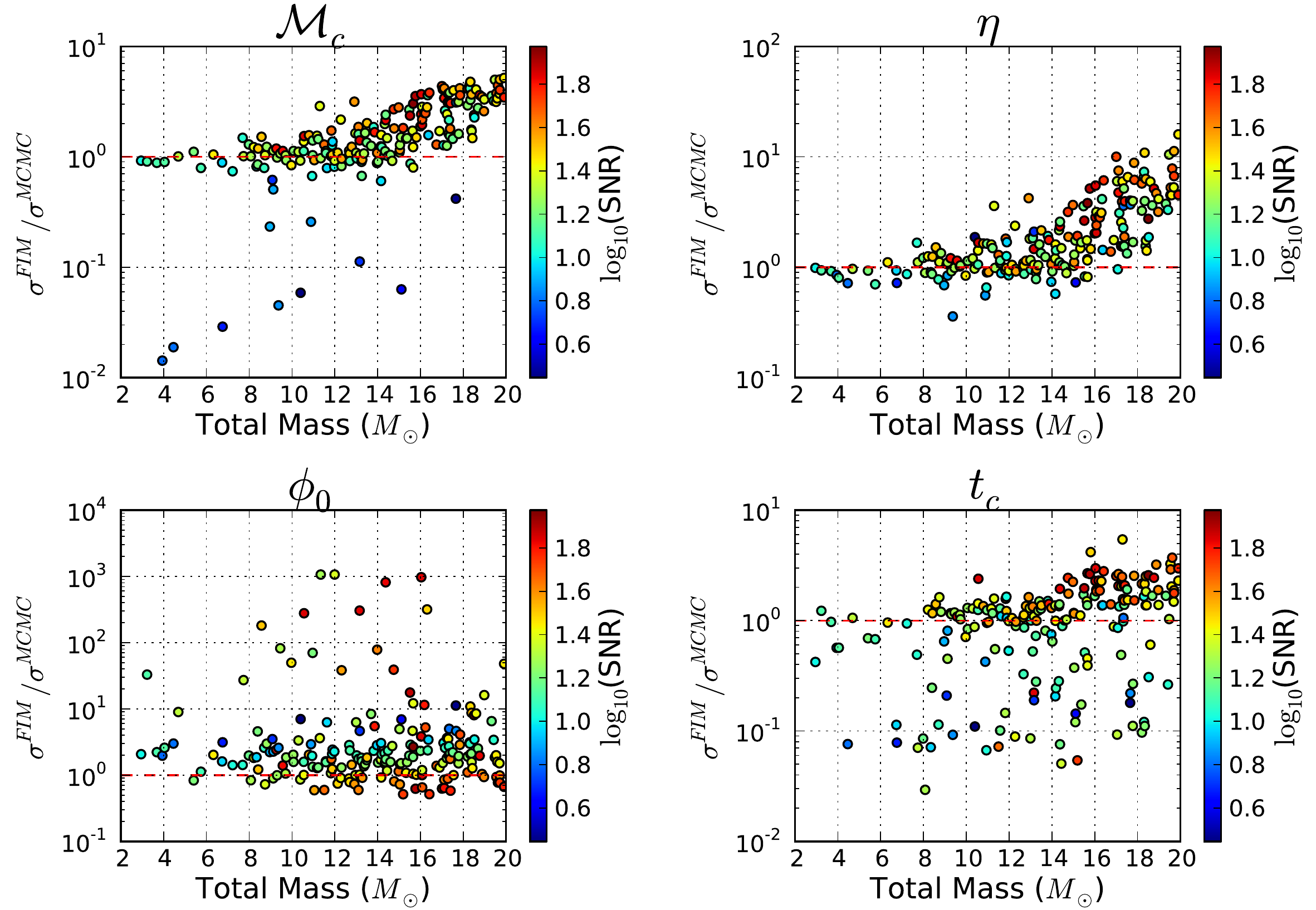}
  \caption{The fractional difference between the marginalized FIM and
    MCMC standard deviations of the intrinsic and phasing parameters
    as a function of total system mass.  The color bar illustrates the
    $\log_{10}(\text{SNR})$ of the injection, while the red line at
    unity represents agreement between the FIM and MCMC errors.  Note
    in particular the systematic divergence at $M > 10M_{\odot}$ in
    the two mass parameters $\mathcal{M}_c$ and $\eta$.  Also note that the
    uncertainties in $\phi_0$ can often be severely overestimated by the
    FIM.}
  \label{fullMismatchFigureIntrinsic}
\end{figure*}

\begin{figure*}[tp!]
  \centering
  \includegraphics[scale=0.74]{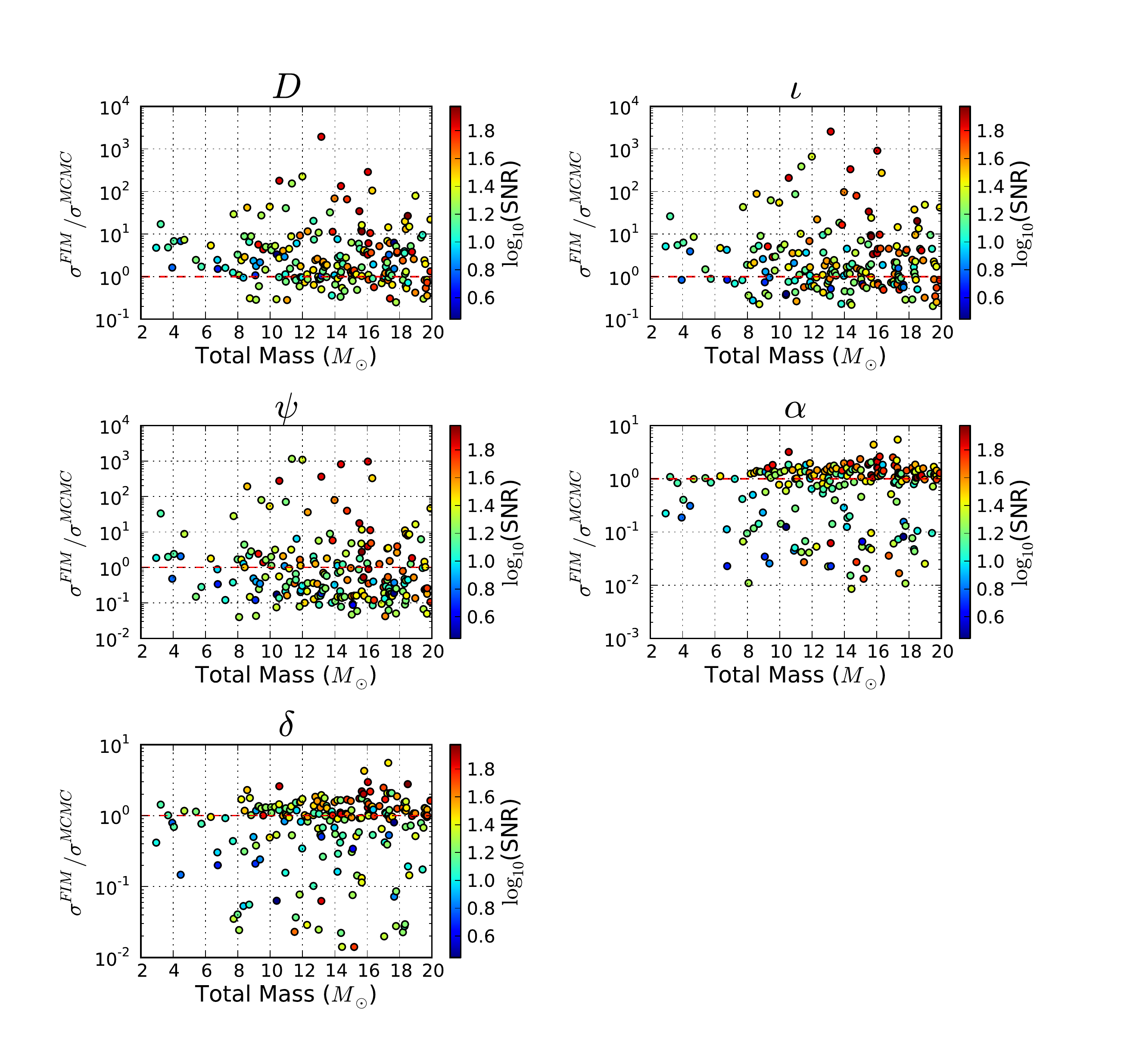}
  \caption{The fractional difference between the marginalized FIM and
    MCMC standard deviations of the extrinsic parameters as a function
    of total system mass.  Unlike the intrinsic parameters
    (Fig. \ref{fullMismatchFigureIntrinsic}), the uncertainty fraction
    does not depend on system mass. }
  \label{fullMismatchFigureExtrinsic}
\end{figure*}

To quantify the difference between the uncertainty estimates on
individual parameters, we use the uncertainty fraction,
\begin{equation}
\Lambda \equiv \frac{\sigma^{\text{FIM}}}{\sigma^{\text{MCMC}}}.
\label{errorFraction}
\end{equation}
where $\sigma^{\text{FIM}}$ is the standard deviation predicted by the
Fisher matrix and $\sigma^{MCMC}$ is the standard deviation of the
posterior distribution. An uncertainty fraction near unity indicates
that the FIM estimate and MCMC standard deviation agree, whereas a
value less than unity indicates the FIM uncertainty is smaller than
the MCMC uncertainty, while values above unity indicate that the MCMC
standard deviation is smaller than the FIM estimate.  

We simulate 200 random, non-spinning systems using the TaylorF2
waveform described in section \ref{waveformSection}.  For the MCMC, we
inject the signals into randomly generated noise realizations using
simulated Initial LIGO Gaussian noise, \eqref{PSD}.  The source distribution, which was then employed as the prior distribution by the MCMC, $p(\partheta)$, is:
\begin{itemize}
\item uniform in component masses in $1M_{\odot} \leq m_1,m_2 \leq 15M_{\odot}$,  
with a total-mass cutoff of $M \leq 20M_{\odot}$;
\item uniform in the logarithm of luminosity distance, $\log(D)$, with a range of
$10\,\text{Mpc} \leq D \leq 40\,\text{Mpc}$; and
\item uniform in all other parameters.
\end{itemize}

We plot the
uncertainty fraction, $\Lambda$, for each of the 9 parameters in our
waveform model.  The results for intrinsic and phasing parameters are shown in
Figure \ref{fullMismatchFigureIntrinsic}.  The results
for the extrinsic parameters are shown in Figure \ref{fullMismatchFigureExtrinsic}.

We focus first on the two mass parameters, $\mathcal{M}_c$ and $\eta$,
as these are the intrinsic parameters of direct physical interest in
most studies.  Immediately, Fig. \ref{fullMismatchFigureIntrinsic}
reveals two distinct features.  First, a sufficiently low SNR will
cause the uncertainty fraction $\Lambda$ to drop significantly below
unity.  In particular, points on the $\mathcal{M}_c$ plot of
Fig. \ref{fullMismatchFigureIntrinsic} below $\rho \approx 8$ show a
mismatch between the two techniques of several orders of magnitude in
some cases.

The second feature is the systematic divergence of $\Lambda$
\textit{above} unity for systems with a high total mass.  As the mass
of the system increases, the MCMC analysis appears to return standard
deviations smaller than those predicted by the FIM.  At high masses
($\sim 20M_{\odot}$), the FIM can over-estimate the MCMC standard
deviations by a factor of 4 in chirp mass, and more than a factor of
10 in $\eta$.  The large (30) number of points above $18 M_{\odot}$,
all with $\Lambda > 1$, suggest the effect is neither a fluke of the
parameter space nor due to a specific realization of the noise, but is
in fact a systematic failure of the FIM to reliably estimate the
posterior variance at high masses.  We will explore the reasons for
this discrepancy in Section \ref{breakdownSection}.

Turning to the extrinsic parameters, there appears to be no distinct
trend similar to that seen in the intrinsic parameters.  For the
errors in $D$, $\iota$, and $\psi$, the FIM can over-estimate the
standard deviations of the MCMC analysis by several orders of
magnitude.  For the luminosity distance, this is most likely due to
the non-uniform shape of the prior probability distribution.  The MCMC
  assumes a prior distance evenly distributed in $\log(D)$.
 For the inclination and wave polarization, the
disparity between the FIM and MCMC results is likely due to how the
two techniques are computing the posterior.  The MCMC is evaluating
the full likelihood surface, in which waveforms at $\iota$ and $\iota
+ \pi$ are identical.  This knowledge restricts the errors returned by
the MCMC to (roughly) their physical limits (which, in the case of
inclination, means the error will always be less than $\pi$ radian).
The FIM, meanwhile, is computed at a local point in parameter space,
and does not have access to the knowledge that the likelihood function
is periodic in certain parameters (see Section \ref{priorsSection}). 
Furthermore, since inclination is
  highly correlated with luminosity distance, the non-uniform prior
  imposed by the MCMC on distance will affect the uncertainties in
  $\iota$ via the cross-correlation.
  
It is surprising that the FIM appears, at first glance, to provide an
adequate error estimate for the sky location of the source.  The
average $\Lambda$ for the sky angles $\alpha$ and $\delta$ is 0.97 and
1.00, respectively.  Since the calculation of the Fisher Matrix
includes the full three-detector network and the geocentric
time-of-arrival difference (folded into $t_c$ in equation
\eqref{phase} for each individual detector site), the FIM has access
to the same amount of local information as the full MCMC.  This does
\emph{not} mean that the FIM standard deviations for a single point
in the sky-position parameter space can be trusted: for a single point
in the sky, several factors (multimodal probability distributions,
non-Gaussian posteriors, etc.) can cause the FIM and MCMC standard
deviations to disagree by up to three orders of magnitude.  Furthermore, we note that the FIM tends to highly underestimate the
error for the low-SNR cases, and slightly overestimate the error for
the high-SNR cases (beginning at about $\rho = 25$).  
While this effect averages to unity for the current study, we note
that this is obviously dependent on the distribution of source SNRs.
As the set of injections were not distributed uniformly in volume, we
expect that this result will \emph{not} hold in general. However, as
an ensemble average, the results indicate that the FIM \emph{can} be
useful for large scale studies comparing the sky-localization
abilities of various detector configurations and designs
(e.g. \cite{Veitch2012}).

\section{Breakdown of the Fisher Matrix}
\label{breakdownSection}

Having demonstrated the existence of a systematic issue with the
Fisher Matrix uncertainty estimates, we now explore potential causes
of the high-mass disagreement between the FIM and MCMC standard
deviations.  We analyze a few of the possible situations that can
arise when using the Fisher Matrix, and attempt to quantify which of
these are causing the discrepancy in the presented results.

\subsection{The Problem of Priors and Restricted Parameters }
\label{priorsSection}

\begin{figure*}[tbp!]
 \centering
 \includegraphics[scale=0.7]{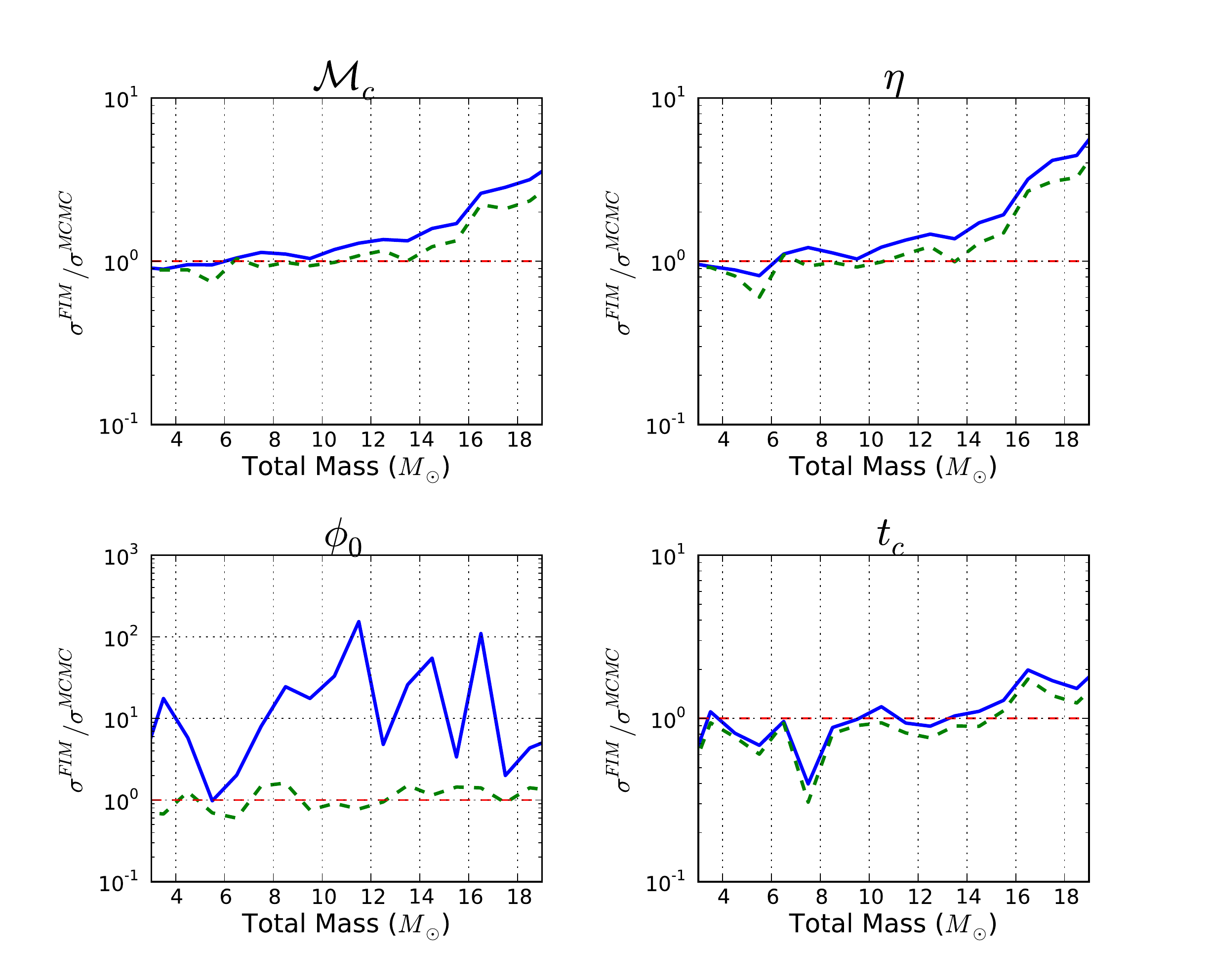}
  \caption{The averaged fractional differences between the FIM and
    MCMC standard deviations, $\Lambda$, for FIM runs with and without
    a Gaussian prior on $\phi_0$ (dashed green and solid blue,
    respectively).  We have averaged the points with a bin size of
    $1M_{\odot}$ in order to better illustrate the difference between
    the two datasets.  Points with SNRs below $\rho = 10$ have been
    excluded. The phase prior clearly lowers the FIM estimates of
    uncertainties on other parameters due to the high correlations;
    however, the effect is only one contribution to the high mass
    divergence of $\sigma_{\text{FIM}}$ and $\sigma_{\text{MCMC}}$.}
 \label{fractionWithPriors}
\end{figure*}

One issue of the Fisher Matrix formalism is the restriction of the
domain of \eqref{FIM} to $\mathbb{R}^n$.  In reality, of course,
several of our parameters in $\partheta$ are angular (e.g. $\phi_0$,
$\psi$), and are restricted to $\mathbb{S}^1$.  For instance, the
coalescence phase, $\phi_0$, can only take values in the range of 0 to
$2\pi$, and usually has a flat posterior.  Since the Fisher
Matrix is attempting to fit a Gaussian
distribution on $\mathbb{R}^1$ to a uniform distribution on
$\mathbb{S}^1$, standard deviations on $\phi_0$ can reach into the thousands
of radians.  Given that the coalescence phase tends to be highly
correlated with the chirp mass (see Table \ref{corrTable}), it is
inevitable that some of the unreasonably large uncertainties
calculated for $\phi_0$ would induce unphysically large deformations
of the $\mathcal{M}_c$ and $\eta$ error ellipses.  Appendix
\ref{app:bp} gives a toy example illustrating how the general effect
of bounded priors is to introduce bias into the estimator, rendering
the unbiased Cramer-Rao bound inappropriate.  

\begin{table}[t!]
\caption{The average normalized cross-correlations between the two
  mass parameters and the two ``junk'' intrinsic parameters.  Note in
  particular the high median correlations with coalescence phase
  $\phi_0$.  This allows unphysical uncertainties in coalescence phase
  (those larger than $2\pi$) to bias mass estimate uncertainties when
  evaluated using the FIM.}
    {\renewcommand{\arraystretch}{1.3} 
\begin{tabular}{lcccc}

\hline\hline Correlation & \vline & Mean & \vline & Median\\

\hline $|C(\mathcal{M}_c ,\phi_0)|$ & \vline & 0.66 &\vline &
0.88\\ $|C(\mathcal{M}_c, t_c)|$ & \vline & 0.72 & \vline &
0.78\\ $|C(\eta, \phi_0)|$ & \vline & 0.69 & \vline & 0.93\\ $|C(\eta,
t_c)|$& \vline & 0.78 & \vline & 0.85\\ \hline\hline

\end{tabular}}
\label{corrTable}
\end{table}

To test the effects of these unphysical standard deviations on our
parameters of interest, we rerun the FIM over the injection points
with a Gaussian prior of width $\sigma_{\phi_0}^{\text{prior}} =
2\pi$.  A Gaussian prior can be easily incorporated into the Fisher
matrix \cite{PoissonWill,Vallisneri}; more complicated priors cannot.
The results for the intrinsic parameters are shown, alongside the flat
prior results, in Fig. \ref{fractionWithPriors}.  Adding a prior
on $\phi_0$ to the FIM analysis decreases, but does not correct, the divergence
between the FIM and MCMC estimates in the high-mass, high-SNR regime. 
 Although a more restrictive
prior might remove the divergence completely (the
$\sigma_{\phi_0}^{\text{prior}} = 2\pi$ prior still allows significant
support outside the $[0,2\pi]$ allowed range for $\phi_0$), there is no physical justification for such a restriction.

However, it is important to point out that $\phi_0 \in [0,2\pi]$
priors are used in MCMC calculations only for the sake of speed.  Very
wide priors could be used instead, and while the marginalized
posterior on the coalescence phase would obviously change, the
marginalized posterior on chirp mass would be unaffected.  That is
because waveforms with $\phi_0 \to \phi_0 + 2n\pi$ are identical and
have identical likelihoods as computed in the MCMC.  Therefore, it is
not obvious that these compact, angular parameters are really to blame for the
apparent violation of the Cramer-Rao bound.

We have only considered a prior on coalescence phase because $\phi_0$
is the only angular parameter in the 9-dimensional model that is
strongly correlated with the two mass parameters.  Priors on extrinsic
parameters, while useful for those parameters, were found to have a
minimal effect on the mass parameters.  Therefore, prior information
on sky position, inclination, distance, etc. does not translate en
masse to the errors on $\mathcal{M}_c$ and $\eta$, since the
cross-correlations between those parameters are relatively
insignificant.

A potentially more problematic prior is the physical boundary at $\eta
= 0.25$ in the mass parameter space.  If, for instance, a signal were
to have injected parameters at $\eta = 0.25$, the standard deviations
returned by the FIM would range to unphysical values of the mass
ratio.  The MCMC errors suffer no such drawback as they are not limited
 to Gaussian-only posteriors.  It is then reasonable to assume that the 
 mass-ratio
cutoff has influenced all of the FIM results quoted in Figures
\ref{fullMismatchFigureIntrinsic} and \ref{fractionWithPriors}.  

As stated above, there is no simple way to include non-Gaussian prior
information in the Fisher-matrix formalism.  However, the majority of
our 200 signals (65\%) were selected with sufficiently
asymmetric mass ratios such that the 1-$\sigma$ surface about the injected values
returned by the FIM did not exceed the physical boundary at $\eta
=0.25$.  We conclude that while an exact prior would affect the
results, the overall trend itself cannot be explained by a prior
boundary; otherwise, some of the high-mass points (those whose
1-$\sigma$ boundaries were completely physical) in the
$\mathcal{M}_c$ and $\eta$ plots of
Fig. \ref{fullMismatchFigureIntrinsic} would lie at unity, which is
not observed.  Additionally, the previous study by Cokelaer
  \cite{cokelaer} attempted to correct this issue by removing the
  boundary at $\eta = 0.25$ in their Monte Carlo parameter estimation.
  Even when $\eta$ was allowed to assume nonphysical values, it was
   found that the FIM still returned uncertainties
  larger than those generated by a full parameter-estimation search.
  
In general, an MCMC analysis will out-perform the Cramer-Rao bound
 when the posterior probability is significantly truncated by \emph{any} boundaries in
 a non-Gaussian prior.   For the initial results 
  reported in section \ref{resultsSection}, the MCMC was performed with a
  maximum component mass of $m_{1,2} \leq 15M_{\odot}$, and a total mass 
  of $m_1 + m_2 \leq 20 M_{\odot}$, in addition to the $\eta \leq 0.25$ restriction.
    Given the population of samples, some of which are injected close to these boundaries, there 
  are some systems whose MCMC posteriors encounter the prior boundaries.  Were
  the Fisher Matrix to be evaluated in component-mass parameters in these regions, the corresponding
  1-$\sigma$ surfaces would also extend beyond these boundaries (a fact which
  is encoded in the 1-$\sigma$ values for $\mathcal{M}_c$ and $\eta$ in this study).  However, 
  when considering all possible prior boundaries, we find that 
  at least $33\%$ of the 27 MCMC results with a total mass greater than $18M_{\odot}$
  do not possess significant probability support near these mass cutoffs.  We conclude that while the choice
  of prior has affected the results quoted here, it cannot completely explain the divergence observed in Figure
   \ref{fullMismatchFigureIntrinsic}.  In Section \ref{biasSection}, we perform an analysis 
   with no prior boundaries in the total or component masses, the results of which solidify this claim.
  
 \subsection{Breakdown of the LSA}
 \label{vallisneriSection}

 Although it does not explain the observed violation of the Cramer-Rao bound, the
 breakdown of the linear-signal approximation at low SNRs can provide a clear example of when the
 FIM results should be treated as suspect.   Vallisneri \cite{Vallisneri} developed a
   criterion to determine if the SNR of a single signal was
   sufficiently loud for the LSA to be considered valid.  By comparing
   the difference between two nearby evaluations of the waveform to the
   linearized shift in the waveform, $\theta^i h_i (\parthetazero)$,
   one can quantify how valid the LSA is for a specific
   $\parthetazero$ and SNR.  In this case, the differences are
   computed between the waveform at the true value and at a random
   point on the 1-$\sigma$ surface, $\Delta h = h(\boldsymbol{\theta_0
     + \sigma})-h(\parthetazero) $).  The overlap of the residuals, $r$, is
   defined as

 \begin{equation}
 \vline ~\log(r)~ \vline \equiv \left< \theta^i h_i - \Delta h ~\vline~
 \theta^j h_j - \Delta h \right> / 2
 \label{valisCriterion}
 \end{equation}

 \noindent By computing this overlap ratio for a large number of points
 distributed over the 1-$\sigma$ surface predicted by the FIM, the
 ratio $\log(r)$ provides an estimate of how linearized the waveform is
 at that point in parameter space.  Furthermore, since the size of the
 1-$\sigma$ surface depends on the signal SNR, \eqref{valisCriterion}
 can be used to determine the linearity of a waveform as a function of
 SNR, and therefore how appropriate the FIM is for a given system.

 We applied the Vallisneri criterion to each of the 200 signals
 presented above, by computing the overlap of 1000 points evenly distributed
 on each 1-$\sigma$ surface.  Following \cite{Vallisneri}, a particular system 
 was considered to be in the linear regime
  if $\vline ~\log(r) ~\vline \leq 0.1$ for 90\% of the
 points.  While the criterion is both intuitive and useful, we found
 that the majority of systems failed the consistency check.
 This held true even for the low-mass points where the FIM and MCMC
 estimates agreed.  We concluded that the criterion provides a sufficient but not
 necessary check of the self-consistency of the Fisher Matrix.

\subsection{The Necessity of Averaging over Many Noise Realizations}
\label{systematicSection}

\begin{figure}[tp!]
  \centering \includegraphics[scale=0.45]{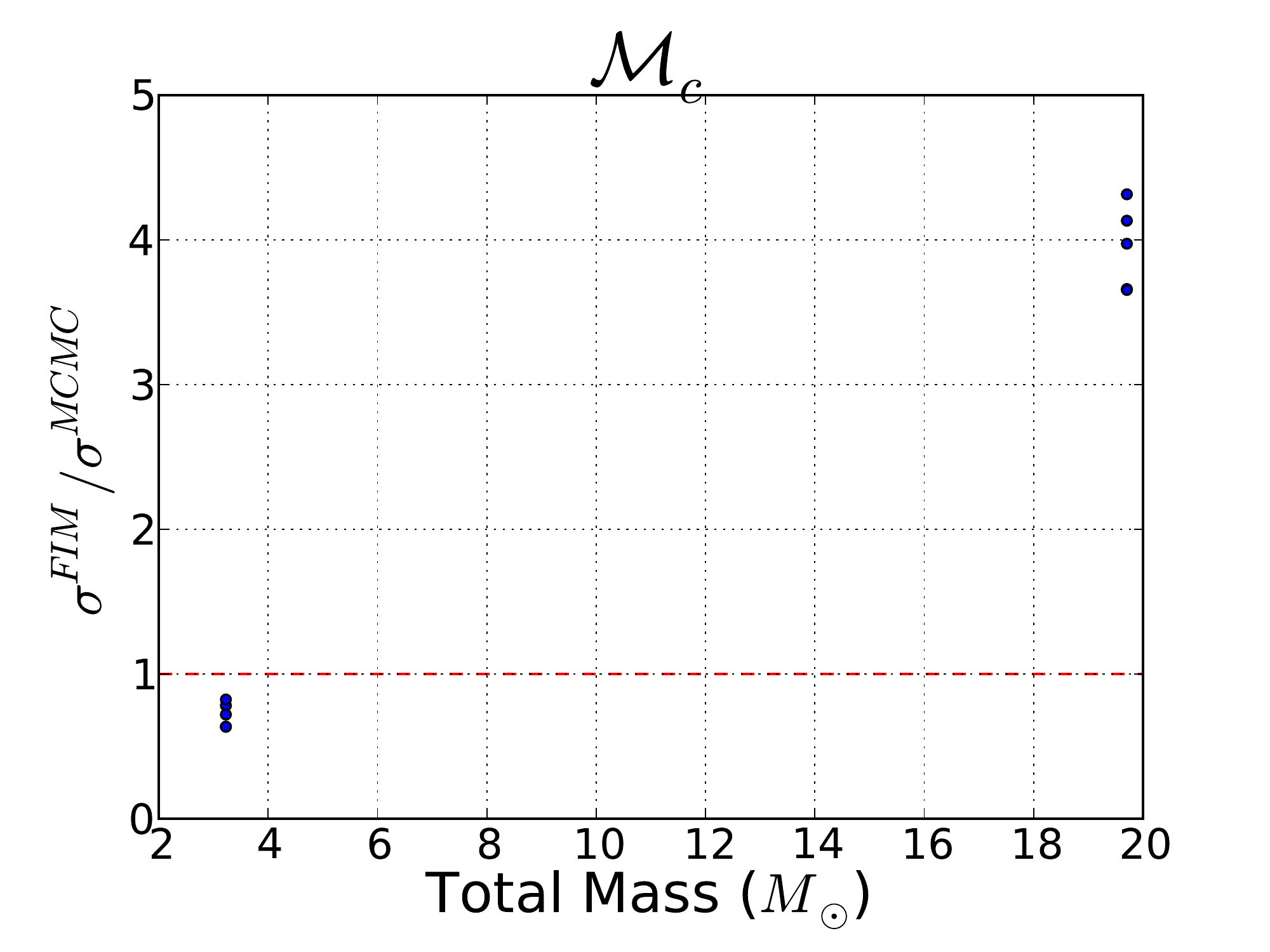}
  \caption{The fractional difference between the FIM and MCMC standard
    deviations of the chirp mass as a function of total system mass
    for two points with total masses $3.2 M_{\odot}$ and $19.7
    M_{\odot}$.  The two systems were selected with parameters at least  1-$\sigma$ away
     from any prior boundary (as measured by the FIM), and with no significant posterior
     support near the same prior boundaries (as measured by the MCMC).  
      The vertical spread represents different MCMC
    standard deviations taken from runs with identical parameters with
    different randomly generated realizations of the detector noise.
    Note how the low-mass points are all
      below unity, suggesting the FIM is providing an adequate lower
      bound, while the high-mass points are all in substantial
    disagreement with the MCMC predictions, even allowing for
     differences from specific noise realizations.}
 \label{MCMCReruns}
\end{figure}

The MCMC and Fisher-matrix analyses address different statistical
ensembles.  The MCMC analysis treats the data as fixed, and the
parameters are allowed to vary; the Fisher analysis treats the
parameters as fixed and the data as random.  Thus, it is expected that
the two methods will not agree for the analysis of any one signal.
However, when an average is taken over many different signal
realizations (parameter choices) and data (noise choices), the results
should be on average consistent with each other in the appropriate
limits.  We believe that we have enough analyses with different noise
and parameter values that the trend towards higher errors at higher
masses observed in Section \ref{resultsSection} is robust.  As a
check, we selected two injections with total masses of $3.2 M_{\odot}$
and $19.7 M_{\odot }$, and ran the MCMC on data corresponding to these
signals with 5 distinct noise realizations.  The results are shown in
Fig. \ref{MCMCReruns}.  As expected, the points do display a spread in
values when the noise realization of the MCMC is changed.  However,
the spread is substantially narrower than the distance above unity for
the high-mass system; we conclude that the upward trend in
Fig. \ref{fullMismatchFigureIntrinsic} is robust.

We have also observed the same trend in a Fisher matrix analysis when
the FIM is computed at the empirical maximum-likelihood parameters
found in the MCMC analysis, rather than at the `true' injected
parameters.  This served to exclude the first source of noise-based covariance (the displacement of the mean)
identified in \eqref{eq:WillMagic}. The results are consistent with the Fisher matrix
analysis at the true parameters $\partheta_0$; we list the
change in the average standard deviations in Table
\ref{injectMaxLTable}.

\begin{table}[t!]
\caption{Mean and median uncertainty
    fractions for the 200 injections, with the FIM evaluated at the
  injected parameters versus the maximum likelihood point of the MCMC.
For the two mass
  parameters, the average change in standard deviations is negligible.
  The only parameters to exhibit a large shift are the mean values for
  $\psi$ and $\phi_0$.  This is due to the presence of outliers, and
  does not appear to affect estimation of the other parameters
  on average. }
  {\renewcommand{\arraystretch}{1.3} 
\begin{tabular}{lcccccccc}

\hline\hline Parameter & \vline & $\Lambda_{\text{Inject}}$ & \vline &
$\Lambda_{\text{MaxL}}$ & \vline & $\Lambda_{\text{Inject}}$ & \vline
& $\Lambda_{\text{MaxL}}$\\ \hline & \vline & \multicolumn{3}{c}{Mean}
& \vline & \multicolumn{3}{c}{Median}\\

\hline $\mathcal{M}_c$ & \vline & 1.84 &\vline & 1.82 &\vline & 1.43
&\vline & 1.39\\ $\eta$ & \vline & 2.43 & \vline & 2.37&\vline & 1.49
&\vline & 1.43\\ $\phi_0$ & \vline & 29.67 & \vline & 211.24 &\vline &
1.96 &\vline & 1.43\\ $t_c$& \vline & 1.20 & \vline & 1.20&\vline &
1.14 &\vline & 1.16 \\ $D$& \vline & 20.65 & \vline & 11.82&\vline &
2.06 &\vline & 0.91\\ $\iota$& \vline & 32.79 & \vline & 43.23&\vline
& 1.67 &\vline & 0.71\\ $\psi$& \vline & 29.57 & \vline &
211.67&\vline & 0.46 &\vline & 0.22\\ $\alpha$& \vline & 0.97 & \vline
& 0.94&\vline & 1.04 &\vline & 0.98\\ $\delta$& \vline & 1.00 & \vline
& 0.97&\vline & 1.00 &\vline & 1.00\\ \hline\hline

\end{tabular}}
\label{injectMaxLTable}
\end{table}

\subsection{Hidden Prior Bias}
\label{biasSection}

\begin{table*}[tbh!]
\caption{ Comparisons of the
    statistical uncertainty between the FIM and MCMC for the limited
  dimensionality, flat prior tests for low-mass and high-mass systems.
  The MCMC errors for the two systems are reported as two separate
  values.  $\sigma^{\text{MCMC}}_{\text{mean}}$ is the frequentist standard
  deviation of the means, calculated by considering the mean of each MCMC posterior as
  the output of an estimator and computing the standard deviation of that estimator.
Mean$(\sigma^{\text{MCMC}})$ is the mean of the Bayesian posterior standard deviations,
computed by averaging the standard deviation of each MCMC posterior. The
  FIM uncertainties $\sigma^{\text{FIM}}$ are also
  presented.  The 2-D problem estimates only the mass parameters,
  $\mathcal{M}_c$ and $\eta$, while the 3-D problem also includes
  $\phi_0$.  In the 2-D problem, the low-mass system obeys the
  Cramer-Rao bound, and the high-mass system is very nearly within the
  Cramer-Rao bound.  In the 3-D problem the low-mass system obeys the
  Cramer-Rao bound, but in the high-mass system the MCMC standard
  deviations are below the Cramer-Rao lower bound. }
    {\renewcommand{\arraystretch}{1.3} 
\begin{tabular}{lcccccccccccccc}

\hline\hline \multicolumn{3}{c}{\textbf{Low Mass ($4.043M_{\odot}$)}}
& \vline & \multicolumn{5}{c}{\textbf{2-D Problem ($\mathcal{M}_c$ and
    $\eta$)}} & \vline & \multicolumn{5}{c}{\textbf{3-D Problem
    ($\mathcal{M}_c$, $\eta$, and $\phi_0$)}} \\

\hline Parameter & \vline & Injected Value & \vline &
$\sigma^{\text{MCMC}}_{\text{mean}}$ & \vline &
Mean$(\sigma^{\text{MCMC}})$ & \vline & $\sigma^{\text{FIM}}$& \vline
& $ \sigma^{\text{MCMC}}_{\text{mean}}$ & \vline &
Mean$(\sigma^{\text{MCMC}})$ & \vline & $\sigma^{\text{FIM}}$\\

\hline $\mathcal{M}_c$ & \vline & 1.650 &\vline & 8.17$\times 10^{-5}$
&\vline & 8.09$\times 10^{-5}$ &\vline & 8.04$\times 10^{-5}$&\vline &
2.03$\times 10^{-4}$ &\vline & 1.64$\times 10^{-4}$ &\vline &
1.62$\times 10^{-4}$\\ 

$\eta$ & \vline & 0.2244 & \vline & 1.97$\times10^{-4}$&\vline &
1.84$\times10^{-4}$ &\vline & 1.83$\times10^{-4}$& \vline &
1.05$\times10^{-3}$&\vline & 8.34$\times10^{-4}$ &\vline &
8.25$\times10^{-4}$\\ 

$\phi_0$ & \vline & 2.38 & \vline & -- &\vline & -- &\vline & -- &
\vline & 0.178 &\vline & 0.142 &\vline & 0.141\\ 

\hline\hline \multicolumn{3}{c}{\textbf{High Mass ($18.20M_{\odot}$)}}
& \vline & \multicolumn{5}{c}{\textbf{2-D Problem ($\mathcal{M}_c$ and
    $\eta$)}} & \vline & \multicolumn{5}{c}{\textbf{3-D Problem
    ($\mathcal{M}_c$, $\eta$, and $\phi_0$)}} \\ 

\hline Parameter & \vline & Injected Value & \vline &
$\sigma^{\text{MCMC}}_{\text{mean}}$ & \vline &
Mean$(\sigma^{\text{MCMC}})$ & \vline & $\sigma^{\text{FIM}}$& \vline
& $ \sigma^{\text{MCMC}}_{\text{mean}}$ & \vline &
Mean$(\sigma^{\text{MCMC}})$ & \vline & $\sigma^{\text{FIM}}$\\ 

\hline $\mathcal{M}_c$ & \vline & 6.257 &\vline & 2.57$\times 10^{-3}$
&\vline & 2.72$\times 10^{-3}$ &\vline & 2.73$\times 10^{-3}$ &\vline
& 5.88$\times 10^{-3}$ &\vline & 5.46$\times 10^{-3}$ &\vline &
7.07$\times 10^{-3}$\\ 

$\eta$ & \vline & 0.169 & \vline & 2.38$\times10^{-4}$&\vline &
2.66$\times10^{-4}$ &\vline & 2.68$\times10^{-4}$& \vline &
1.34$\times10^{-3}$&\vline & 1.22$\times10^{-3}$ &\vline &
1.66$\times10^{-3}$\\ 

$\phi_0$ & \vline & 5.32 & \vline & -- &\vline & -- &\vline & -- &
\vline & 0.147 &\vline & 0.164 &\vline & 0.194\\ 

\hline\hline

\end{tabular}}
\label{2D3DRerunsTable}
\end{table*}

The Cramer-Rao bound as stated applies to only unbiased
estimators of a parameter.  But the bound in Eq.~\eqref{FIM}
only incorporates information from the likelihood, while the MCMC analysis
uses the posterior, which includes a prior distribution,
$p(\partheta)$.

By default, the uniform priors on the component masses \texttt{lalinference\_mcmc} employs 
yield highly
non-flat priors in $\mathcal{M}_c$ and $\eta$.  In effect, our choice
of prior may provide a biased estimator with which we could surpass
the Cramer-Rao bound.  As a final test of the results, we reduce the
comparison to the simplest possible case.  We choose a low-mass and
high-mass injection from the 200 injections.  Both systems were selected with
a sufficiently asymmetric mass ratio to ensure no interference from
the physical cutoff at $\eta=0.25$.  To remove any possible bias and
correlations, we employ uniform priors in the MCMC
\begin{align}
p(\mathcal{M}_c) &= 1 ~~~~ 0.8M_{\odot} \leq \mathcal{M}_c \leq
10M_{\odot}\\ &= 0 ~~~~\text{otherwise}\nonumber\\ p(\eta) &= 1 ~~~~
0.05 \leq \eta \leq 0.25\\ &=0 ~~~~ \text{otherwise}\nonumber
\end{align}
and reduce the MCMC to a two-dimensional space over only
the mass parameters $\mathcal{M}_c$ and $\eta$, with all
 other parameters held fixed at their
true values.  This choice of priors also eliminates the issue of boundaries in total and component
masses discussed in Section \ref{priorsSection}.  We then performed an MCMC on the two injections into 50
different realizations of Gaussian noise, thus enabling a valid comparison
between the spread of the MCMC estimators and the standard deviations
returned by the FIM.  For the reduced dimensionality runs, we find
near perfect agreement between the standard deviation of the MCMC
estimates and the Cramer-Rao bound for both mass cases.  We also find
support for the claim that the average of the uncertainty estimates of
the individual MCMC experiments equals the standard deviation of
posterior means over multiple noise realizations.  
%\ilya{[Is it
 %   worthwhile to try to prove this claim more formally?  BTW, should
 %   I be worried that the 3-D low mass results violate this claim - by
 %   more than an order of magnitude in one case?]}\carl{[I'm somewhat
 %   worried now that I've corrected the 2D high mass results.  I'm
 %   going to proceed without a proof and work on it later.  If I can't
 %   come up with something worthwhile I'll just remove the claim.]}
See Table \ref{2D3DRerunsTable}.

As a test of our hypothesis in section \ref{priorsSection}, we then
perform the same analysis over a three-dimensional parameter space
consisting of the two mass parameters and the chirp phase
($\mathcal{M}_c$, $\eta$, and $\phi_0$).  In this case, we find that
the MCMC matches the uncertainty estimates predicted by the FIM for
the low-mass system.  For the high mass system, the MCMC
still produces standard deviations which significantly surpass the
Cramer-Rao bound in all three parameters when $\phi_0$ is included
as a free parameter.  The results are illustrated in Table
\ref{2D3DRerunsTable}.  Notice that for the 3-D case, the uncertainties in $\phi_0$ for
  both the low and high mass systems are substantially less than
  $2\pi$.  This seems to suggest that the $2\pi$ domain of
  $\phi_0$ (and indeed all other angular parameters) cannot be
  responsible for the breakdown of the Cramer-Rao bound in this
  context.
When one adds the coalescence phase parameter into the
analysis, the FIM begins to noticeably fail in the high-mass regime.  
To test whether this effect is due to bias in the $\mathcal{M}_c$ estimator, 
we use the biased form of the
Cramer-Rao bound given in \cite{Vallisneri}:
\begin{equation}
\Sigma^{\text{biased}}_{il} \equiv (\delta_{im} + \partial_m b_i
(\parthetazero))\Sigma^{CR}_{mj}(\delta_{jl} +
\partial_{j}b_{l}(\parthetazero))
\label{biasedCramerRao}
\end{equation}
where $b_{i}(\parthetazero)$ is the bias of the estimator in the
$i^{th}$ parameter, evaluated at the true parameters (see Appendix
\ref{app:cr} for a derivation of the biased Cramer-Rao bound).  In
order for any bias from the boundary in $\phi_0$ to influence the
bound on $\mathcal{M}_c$, for instance, the MCMC estimator would need
to have non-zero first derivatives of the chirp mass bias
$b_{\mathcal{M}_c}(\parthetazero)$.  Since the recovery of the
posterior in the 2-dimensional case appears consistent with the
Cramer-Rao bound, it is reasonable to assume that the two mass
parameters are individually unbiased, such that
$\partial_{\mathcal{M}_c}b_{\mathcal{M}_c}(\parthetazero) =
\partial_{\eta}b_{\mathcal{M}_c}(\parthetazero) = 0$.  Therefore, the
only source of chirp mass bias must arise from the
$\partial_{\phi_0}b_{\mathcal{M}_c}(\parthetazero)$ term.  Under that
assumption, along with the symmetry of $\Sigma$ and equation
\eqref{biasedCramerRao}, the biased Cramer-Rao bound on
$\mathcal{M}_c$ in the 3-dimensional model becomes
\begin{align}
\Sigma^{biased}_{\mathcal{M}_c \mathcal{M}_c} =&
\Sigma^{CR}_{\mathcal{M}_c \mathcal{M}_c} + 2
\Sigma^{CR}_{\mathcal{M}_c \phi_0}
\partial_{\phi_0}b_{\mathcal{M}_c}(\parthetazero)
\nonumber\\ &+ \Sigma^{CR}_{\phi_0
  \phi_0}(\partial_{\phi_0}b_{\mathcal{M}_c}(\parthetazero))^2
\label{biasedChirpMassError}
\end{align}
We can approximate the derivative as
\begin{equation}
\partial_{\phi_0}b_{\mathcal{M}_c}(\parthetazero) \rightarrow
\frac{b_{\mathcal{M}_c}(\phi_0)-b_{\mathcal{M}_c}(\phi_0-\sigma^{FIM}_{\phi_0})}{\sigma^{FIM}_{\phi_0}}
\end{equation} 
using the 1-$\sigma$ FIM errors on $\phi_0$ as a natural derivative
length scale for the problem.  We calculate $b_{\mathcal{M}_c}$ using the mean
of the MCMC posteriors as our estimator.  $b_{\mathcal{M}_c}(\phi_0-\sigma^{FIM}_{\phi_0})$ 
is calculated similarly by performing a second set of 50 MCMC runs with parameters displaced
in $\phi_0$ by $\sigma^{FIM}_{\phi_0}$.  Performing this test, we find that the
chirp mass estimator is not significantly biased by our choice of $\phi_0$
(Table \ref{biasedUnbiasedChirpMassErrors}).  We conclude that the MCMC
estimator is sufficiently unbiased.

%\sout{Based on the results in this section, particularly the re-appearance
%of the Cramer-Rao violation when the parameter space is expanded from
%masses to masses and phase, we expect that the cause is as noted in
%\S~\ref{priorsSection}.  The Fisher matrix does not properly account
%for the prior restriction of the chirp phase to the range $[0, 2\pi]$,
%and the large uncertainties it therefore predicts broaden the mass
%uncertainties artificially due to the correlation among these
%parameters.}

Previous studies have observed similar ``violations'' of the
Cramer-Rao bound \cite{CornishLISA,cokelaer}, as well as a
similar agreement between the FIM and the likelihood for a two-dimensional model
\cite{Keppel13}.  Furthermore, 
Vallisneri \cite{Vallisneri} and Cutler \& Flanagan \cite{CutlerFlanagan} point out that the Cramer-Rao bound
is often unhelpful in practice due to the effects of estimator
bias.  We have speculated on several possible explanations for the
apparent breakdown of the Cramer-Rao bound.  However, despite
exploring a variety of possibilities, we have not been able to find a
completely convincing explanation for this behavior.

\begin{table}[t!]
\caption{$\mathcal{M}_c$ errors for the high-mass, 3-dimensional
  system (bottom right column of Table \ref{2D3DRerunsTable}), but
  employing \eqref{biasedCramerRao} to correct the bias present in the maximum likelihood
  estimator from the compact domain of $\phi_0$.  Although the biased
  Cramer-Rao bound ($\sigma^{\text{FIM}}_{\text{biased}}$) is smaller
  than its unbiased counterpart
  ($\sigma^{\text{FIM}}_{\text{unbiased}})$, it is still larger than
  the MCMC standard deviations.}
    {\renewcommand{\arraystretch}{1.3} 
\begin{tabular}{lcc}

\hline\hline Error Type & \vline & $\sigma_{\mathcal{M}_c}$ ($\times
10^{-3}$ $M_{\odot}$)\\

\hline $\sigma^{\text{MCMC}}_{\text{mean}}$ & \vline &
5.88\\ $\sigma^{\text{FIM}}_{\text{unbiased}}$ & \vline & 7.07
\\ $\sigma^{\text{FIM}}_{\text{biased}}$ & \vline & 6.81\\

\hline\hline

\end{tabular}}
\label{biasedUnbiasedChirpMassErrors}
\end{table}

\section{Conclusion}
\label{conclusionSection}

In this paper we compared the parameter estimation capabilities of one
of the Bayesian codes used by the LIGO Scientific Collaboration,
\texttt{lalinference\_mcmc}, to the theoretical estimates provided by
the Fisher Information Matrix.  The purpose was to compare the
effectiveness of previous FIM predictions to the accuracy achievable
with real parameter estimation techniques that will be employed once
the Advanced LIGO/VIRGO network begins to regularly detect signals from compact
binary coalescense.
  Many studies have used the Fisher information to
describe the science capabilities of advanced detectors; our analysis
provides a route to understanding what these studies imply about the
actual parameter estimation uncertainties when making inference on signals from compact binary coalescences.

We found two distinct effects in the uncertainties of the two mass parameters. For low-SNR
signals, the FIM can produce standard deviations several orders of
magnitude smaller those realizable by an MCMC search.  This is
expected, since the Fisher Matrix as an LSA approximation is based on an expansion in
$1/\text{SNR}$, and requires a sufficiently loud source for the
linear-signal approximation to be valid. However, we also discovered a
systematic divergence between the FIM and MCMC uncertainties for
signals with a total mass above $\sim10_{\odot}$.  Beyond
$16M_{\odot}$ for $\mathcal{M}_c$ and $18M_{\odot}$ for $\eta$, all of
our injected signals were recovered by the MCMC with tighter uncertainties
than the FIM would suggest possible.  This effect was
noted for signals with distinct noise realizations, and with a
wide range of ``reasonable'' SNRs from $\rho \approx 10$ to $\rho
\approx 100$.  

We checked several possible causes for this
discrepancy, including correlations with unconstrained parameters
(e.g. $\phi_0$), the displacement of the MCMC maximum likelihood from the
injected parameters, and possible bias from our choice of 
priors.  We found a slight improvement in FIM estimates by
adding a constraint on $\phi_0$, but even with this (weak) constraint the
overall effect persisted.  We conclude that the breakdown is a
systematic error potentially affecting all FIM predictions, or at least those using
frequency-domain waveforms above $10M_{\odot}$.  This calls into
question several previous parameter-estimation predictions in the
literature \cite[e.g.,][]{ArunPE,PoissonWill,RodriguezIMRI}.  A full parameter
estimation study (currently underway) will be required to determine
the true capabilities of Advanced LIGO/VIRGO, particularly when
constraining the masses of binary black hole systems.  It should also
be noted that, while we employed the Initial LIGO noise curve here
(for computational convenience in the MCMC analysis), the systematic
nature of the trend indicates a potential issue with any known
implementation of the Fisher Matrix analysis
\cite[e.g.,][]{CornishLISA}, and should lend caution to all future
studies.

We also found that the FIM estimates of the extrinsic parameters ($D$,
$\iota$, $\psi$, $\alpha$, $\delta$) were highly variable.  For the
luminosity distance, inclination, and wave polarization, the Fisher
Matrix standard deviations were found to vary wildly relative to those
returned by the MCMC (with the ratios of standard deviations,
$\Lambda$, varying by as much as 4 orders of magnitude).  We
also found that the FIM and MCMC uncertainties on sky-location angles,
$\alpha$ and $\delta$, disagreed by over 2 orders of magnitude in some
cases; we also found that the $\Lambda$ values for the sky-location parameters
($\alpha$ and $\delta$) averaged to 1 over the 200 injected signals.  However, this
result may depend on the injection distribution, and should not be trusted in general.

New techniques are being developed which explore
the parameter space more fully than the FIM, but without the
complications of a full MCMC.  By locally mapping the likelihood
surface, these techniques (such as the exact mapping of the maximum
likelihood estimator \cite{VallisMapping}, or the fitting of an
``Effective Fisher Matrix'' via the local ambiguity function
\cite{EffectiveFisher}), may avoid, or at least make apparent, the
pitfalls we have outlined in this study.  

We must still stress that any FIM study performed without these checks
of the likelihood surface should \emph{not} be trusted blindly.
It is entirely likely that the Fisher Information Matrix
estimates could be failing, and that the failure is not producing an
overly conservative lower-bound on the standard deviation, as has
commonly been assumed, but is producing errors substantially
\emph{worse} than those realizable in practice.  For cases such as
these, one must employ a more robust technique which fully
explores the posterior probability surface.

\begin{acknowledgments}
We thank Michele Vallisneri, Patrick Brady, Frank Ohme, Vicky
Kalogera, Cole Miller, Thomas Dent, Drew Keppel, 
and Alberto Vecchio for useful discussions.  
CR and BF were supported by NSF GRFP Fellowships, award DGE-0824162. 
WF and IM are grateful to to the hospitality of the Kavli Institute for Theoretical Physics, where part of this work was carried out; it was supported in part by the National Science Foundation
under Grant No. NSF PHY11-25915.
\end{acknowledgments}

\bibliography{paper}{}

\appendix

\section{The Cramer-Rao Bound}
\label{app:cr}

The Cramer-Rao bound limits the variance of an unbiased estimator of a
function on a probability space obtained from a finite set of samples
from this space.  Let $d$ be a set of samples drawn according to a
parameterized probability distribution, $p(d | \theta)$:
\begin{equation}
  d \sim p(d | \theta).
\end{equation}
We want to estimate some function of $\theta$, $\psi(\theta)$, using a
function of the data, $d$,
\begin{equation}
  \psi(\theta) \simeq T(d).
\end{equation}
We will assume that the estimator $T$ is unbiased, that is 
\begin{equation}
  \ave{p(d|\theta)}{T(d)} \equiv \int dd\, p(d|\theta) T(d) = \psi(\theta).
\end{equation}

The quantity 
\begin{equation}
  s(d, \theta) \equiv \frac{\partial}{\partial\theta} \ln p(d|\theta)
\end{equation}
is known as the \emph{score}.  As long as we can interchange the order
of integration over $d$ and differentiation with respect to $\theta$,
we can show that the expected value of the score is zero.  We can
interchange integration and differentiation when
\begin{enumerate}
\item The allowed range of $d$ does not depend on $\theta$ and
\item Both $p(d|\theta)$ and $\partial p/\partial \theta$ are
  continuous and
\item If the allowed range of $d$ is infinite (i.e.\ if the integral
  is improper), then all integrals must converge uniformly.
\end{enumerate}
Under these conditions, we have $\ave{p(d|\theta)}{s(\theta)} = 0$ by
the following:
\begin{eqnarray}
  \label{eq:mean-score}
  \ave{p(d|\theta)}{s(d, \theta)} & = & \int dd\, p(d|\theta)
  \frac{\partial}{\partial \theta} \ln p(d|\theta) \nonumber \\ & = &
  \int dd \, \frac{\partial}{\partial \theta} p(d|\theta) \nonumber
  \\ & = & \frac{\partial}{\partial \theta} \int dd\, p(d|\theta)
  \nonumber \\ & = & 0.
\end{eqnarray}
We will assume that conditions 1, 2, and 3 hold unless otherwise
noted.

Consider the covariance of the estimator $T$ and the score:
\begin{multline}
  \cov(T, s) = \\ \ave{p(d|\theta)}{\left( T(d) - \ave{p(d'|\theta)}{T(d')}\right) \left(s(d,\theta) - \ave{p(d'|\theta)}{s(d',\theta)} \right)}
\end{multline}
Since $\ave{p(d|\theta)}{s(d,\theta)} = 0$, this reduces to
\begin{eqnarray}
  \cov(T, s) & = & \ave{p(d|\theta)}{T(d) s(d,\theta)} \nonumber \\
  & = & \int dd\, p(d|\theta) T(d) \frac{\partial}{\partial \theta} \ln p(d|\theta) \nonumber \\
  & = & \int dd\, T(d) \frac{\partial}{\partial \theta} p(d|\theta) \nonumber \\
  & = & \frac{\partial}{\partial \theta} \int dd\, T(d) p(d|\theta) \nonumber \\
  & = & \frac{\partial}{\partial \theta} \psi(\theta). \label{eq:cov-Ts}
\end{eqnarray}

The Cauchy-Schwarz inequality requires
\begin{equation}
  \label{eq:cauchy-schwarz}
\cov(T,T) \cov(s,s) \geq \left(\cov(T,s)\right)^2,
\end{equation}
so 
\begin{equation}
  \label{eq:cr-bound}
  \cov(T,T) \geq \frac{\left( \partial \psi/\partial \theta \right)^2}{I(\theta)},
\end{equation}
where the Fisher information, $I(\theta)$, is defined to be the
covariance of the score:
\begin{equation}
  I(\theta) = \ave{p(d|\theta)}{\left(s(d,\theta)\right)^2} = - \ave{p(d|\theta)}{\frac{\partial^2}{\partial \theta^2} \ln p(d|\theta)}.
\end{equation}
Eq.~\eqref{eq:cr-bound} is the Cramer-Rao bound on the variance of the
estimator $T$.  

\subsection{Biased Estimators}

In the event that the estimator, $T(d)$, is biased, so that 
\begin{equation}
  \ave{p(d|\theta)}{T(d)} = \psi(\theta) + b(\theta),
\end{equation} 
the Cramer-Rao bound is modified to 
\begin{equation}
  \label{eq:cr-bound-biased}
  \cov(T,T) \geq \frac{\left( \partial \psi/\partial \theta +
    \partial b/\partial \theta \right)^2}{I(\theta)}.
\end{equation}
(See Eq.~\eqref{eq:cov-Ts}.)

\section{Bounded Data}
\label{app:bd}

\begin{figure}
  \includegraphics{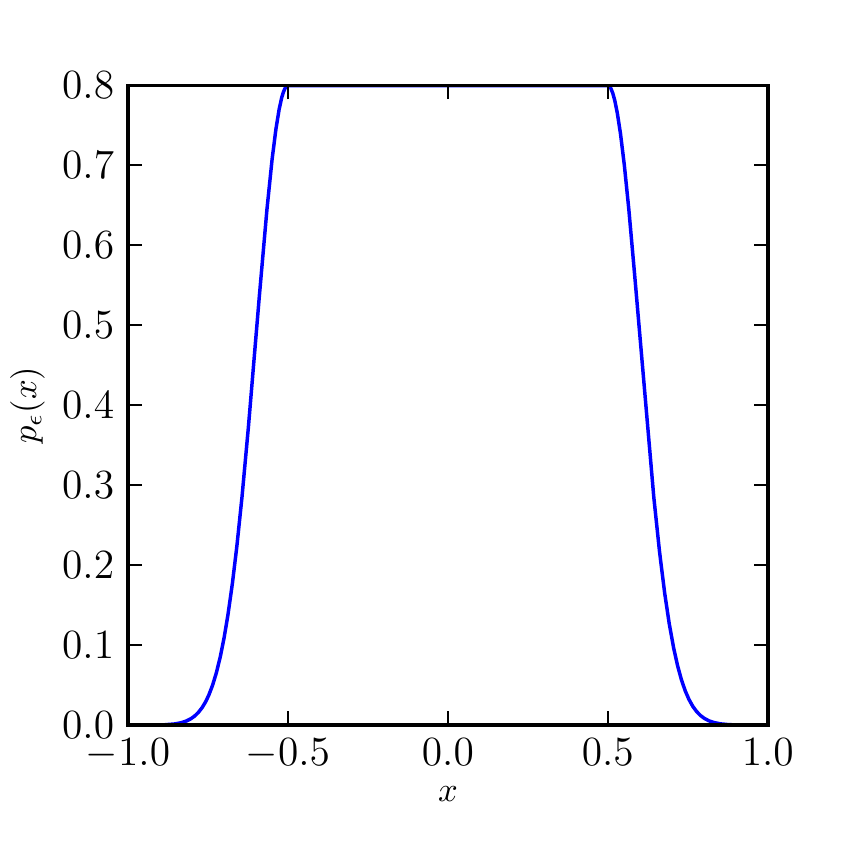}
  \caption{\label{fig:pbounded} The probability distribution
    $p_\epsilon(x)$ defined in Eq.~\eqref{eq:pbounded} with $\epsilon
    = 0.1$}
\end{figure}

Consider the probability distribution defined by 
\begin{equation}
  \label{eq:pbounded}
  p_\epsilon(x) = \frac{1}{1 + \sqrt{2 \pi}\epsilon} \begin{cases}
    \exp\left( -\frac{\left(x + \frac{1}{2}\right)^2}{2 \epsilon^2}
    \right) & x < -\frac{1}{2} \\
    1 & -\frac{1}{2} < x < \frac{1}{2} \\
    \exp\left( -\frac{\left(x - \frac{1}{2}\right)^2}{2 \epsilon^2} \right) &
    x > \frac{1}{2}
  \end{cases}
\end{equation}
A plot of this function appears in Figure \ref{fig:pbounded}.  As
$\epsilon \to 0$, the distribution goes over to a bounded, ``top-hat''
distribution between $-1/2$ and $1/2$, but the distribution is
everywhere (twice) differentiable and nowehere zero for $\epsilon >
0$.  This is required to satisfy conditions 1, 2, and 3 above.

Consider data produced from a parameter $\theta$ with noise drawn from
$p_\epsilon$:
\begin{equation}
  d = \theta + X,
\end{equation}
with 
\begin{equation}
  X \sim p_\epsilon(X).
\end{equation}
As $\epsilon \to 0$, the data concentrate in the bounded range $\theta
\pm 1/2$.  The likelihood for these data is
\begin{equation}
  p(d | \theta) = p_\epsilon(d - \theta).
\end{equation}
Because our distribution is everywhere twice differentiable, we can
compute the Fisher information.  A little algebra reveals 
\begin{equation}
  I(\theta) = \frac{\sqrt{2\pi}}{\epsilon} \frac{1}{1 + \sqrt{2 \pi} \epsilon},
\end{equation}
so the Cramer-Rao bound for estimators, $T(d)$, of $\psi(\theta)$ is
\begin{equation}
  \cov(T,T) \geq \frac{\epsilon}{\sqrt{2\pi}} \left(1 + \sqrt{2 \pi} \epsilon\right)\left( \partial \psi/\partial \theta \right)^2.
\end{equation}
As $\epsilon \to 0$, the Cramer-Rao bound for $T$ becomes trivial.  In
other words, when the data generating distribution has a hard boundary
the Fisher information becomes infinite and the Cramer-Rao bound
becomes trivial.

\section{Bounded Parameter}
\label{app:bp}

Suppose we generate data from the following random process:
\begin{equation}
  d = \theta + n,
\end{equation}
with 
\begin{equation}
  n \sim N(0, \sigma),
\end{equation}
where $N$ is the normal distribution of width $\sigma$, and further suppose we 
have good prior information on $\theta$, that
is:
\begin{equation}
  p(\theta) = U\left(-\epsilon, \epsilon\right),
\end{equation}
where $U$ is the uniform distribution.  We wish to infer the value of $\theta$
 using the data set $d$.  The
Bayesian posterior for $\theta$ after $N$ observations of $d$ is given
by
\begin{equation}
  \label{eq:posterior}
  p(\theta | d) = \begin{cases}
    \alpha \exp\left( - \sum_{i = 1}^N \frac{\left( d_i -
      \theta\right)^2}{2 \sigma^2} \right) \frac{1}{2\epsilon} &
    -\epsilon < \theta < \epsilon \\
    0 & \mathrm{otherwise}
    \end{cases},
\end{equation}
where $\alpha$ is a normalizing constant.  Suppose we want to estimate
$\theta$ using the posterior mean:
\begin{equation}
  \theta \simeq T(d) = \int d\theta \, \theta p(\theta | d).
\end{equation}
After some algebra, we have 
\begin{equation}
  \label{eq:mean-theta-estimator}
  T(d) = \frac{D}{N} - \sqrt{\frac{2}{\pi}} \frac{\sigma}{\sqrt{N}} \frac{\exp\left(-
    \frac{\left( D - N\epsilon\right)^2}{2 N \sigma^2}\right) -
    \exp\left( -\frac{\left(D + N \epsilon\right)^2}{2 N \sigma^2}
    \right)}{\erf\left(\frac{D + N\epsilon}{\sqrt{2N}\sigma}\right) -
    \erf\left(\frac{D - N\epsilon}{\sqrt{2N}\sigma} \right)},
\end{equation}
where
\begin{equation}
  D \equiv \sum_{i=1}^N d_i.
\end{equation}
Because 
\begin{equation}
  \ave{p(d|\theta)}{\frac{D}{N}} = \theta,
\end{equation}
the estimator $T(d)$ is biased.  As one might expect, when 
\begin{equation}
  \epsilon \gg \frac{\sigma}{\sqrt{N}},
\end{equation}
the second term in Eq.~\eqref{eq:mean-theta-estimator} is small, and
the estimator is approximately unbiased; in this limit, the width of
the prior, $2\epsilon$, is large compared to the width of the
posterior, which is approximately $\sigma/\sqrt{N}$.  In the other
limit, when 
\begin{equation}
\epsilon \ll \frac{\sigma}{\sqrt{N}},
\end{equation}
we have
\begin{equation}
  T(d) = \frac{D \epsilon^2}{3 \sigma^2} + \order{\frac{\epsilon^4}{\sigma^4}},
\end{equation}
whence 
\begin{equation}
  \ave{p(d|\theta)}{T(d)} \simeq \frac{N \theta \epsilon^2}{3\sigma^2}.
\end{equation}
In this case, the bias is 
\begin{equation}
  b(\theta) = \ave{p(d|\theta)}{T(d)} - \theta \simeq \theta
  \left(\frac{N\epsilon^2}{3\sigma^2} - 1\right),
\end{equation}
which is very significant (the estimator clusters near zero, no matter
the value of $\theta$ because of the symmetry of the prior).  The
variance of the estimator $T$ in this limit is
\begin{equation}
  \label{eq:narrow-T-variance}
  \cov(T,T) \simeq \frac{\epsilon^4}{9 \sigma^4} \cov(D,D) = N
  \frac{\epsilon^4}{9 \sigma^2}.
\end{equation} 

Note that $\epsilon^2/3$ is the variance of the prior distribution, so
the variance of our estimator is much smaller than that of the
prior \footnote{Be careful! The variance of the prior is an integral
  over the parameter $\theta$, while the variance of the estimator refers
  to the variance at \emph{fixed} $\theta$ under repeated noise
  realizations.}.  Note that the posterior mean is not a particularly useful estimator when the prior is so narrow, since the posterior is dominated by the prior rather than the data, and is roughly uniform in $[-\epsilon, \epsilon]$ while vanishing outside this interval.  While the posterior mean is always close to zero, this carries little information about the true parameter value beyond what is contained in the prior.

The Fisher information for our likelihood function is 
\begin{equation}
  I(\theta) = \frac{N}{\sigma^2}.
\end{equation} 
In both the wide-prior ($\epsilon \gg \sigma/\sqrt{N}$) and
narrow-prior ($\epsilon \ll \sigma/\sqrt{N}$) cases, the Cramer-Rao
bound is satisfied.  For the wide prior case, we have 
\begin{equation}
  T(d) \simeq \frac{D}{N}, 
\end{equation}
whence
\begin{equation}
  \cov(T,T) = \frac{\sigma^2}{N} \geq \frac{1}{I(\theta)} = \frac{\sigma^2}{N}.
\end{equation}
In the wide-prior limit, we achieve the Cramer-Rao bound! 
For the narrow-prior case, we have significant bias.  The numerator of
the Cramer-Rao bound, Eq.~\eqref{eq:cr-bound}, is
\begin{equation}
  \frac{\partial \psi}{\partial \theta} + \frac{\partial b}{\partial
    \theta} \simeq 1 + \frac{N \epsilon^2}{3\sigma^2} - 1 = \frac{N\epsilon^2}{3\sigma^2}.
\end{equation} 
The full Cramer-Rao bound is 
\begin{equation}
  \cov(T,T) \geq \frac{\left(\partial \psi/\partial \theta + \partial
    b/\partial \theta\right)^2}{I(\theta)} = N \frac{\epsilon^4}{9 \sigma^2}.
\end{equation}
Comparing with Eq.~\eqref{eq:narrow-T-variance}, we see that we
achieve the Cramer-Rao bound in the narrow-prior limit, too! 

As an aside, in the narrow prior limit, the varaince of the posterior,
Eq.~\eqref{eq:posterior}, is 
\begin{equation}
  \int d\theta \, \left( \theta - \ave{p(\theta'|d)}{\theta'}
  \right)^2 p(\theta|d) = \frac{\epsilon^2}{3} + \order{\epsilon^4},
\end{equation}
which is much larger than the variance of the estimator (since the
estimator clusters near zero, while the posterior is approximately the
prior in the narrow-prior limit).  

Assuming that the prior is correct (that is, that the true value of
$\theta$, denoted by $\theta_0$, is between $\pm \epsilon$), an
estimate based on the mean of the posterior exhibits lower standard
error than an estimate based on the sample mean.  The standard error
for an estimator is defined as 
\begin{align}
  \se \left( T(d) \right) &\equiv \sqrt{\ave{p(d|\theta_0)}{\left( T(d) - \theta_0
    \right)^2}} \\ &= \sqrt{\cov(T,T) + b^2(\theta_0)}.
\end{align}
When the priors are wide, the mean of the posterior is approximately
equal to the sample mean, and the standard error is then the sample
variance; when the priors are narrow, the posterior mean is nearly 0,
which is always within $\pm \epsilon$ of the true value of $\theta$,
so the standard error is $\sim \epsilon$, which is much smaller than
the variance of the sample mean.  In both cases, the effect of the
prior is to produce a more accurate (but biased!) estimate of the
parameter than can be achieved using the sample mean alone.
 
  \pagebreak
\end{document}